\documentclass[]{elsart}

\usepackage[toc,page]{appendix}

\usepackage{graphicx}

\usepackage{amssymb}

\begin{document}

\begin{frontmatter}



\title{A numerical retro-action model relates rocky coast erosion to percolation theory}


\author{Andrea Baldassarri$^{1,2}$, Bernard Sapoval$^{2,3}$ \and Simon
F\'elix$^{2}$}

\address{$^1$ Dipartimento di Fisica, Universit\`a di Roma ``La
Sapienza'', P.le Aldo Moro 2, 00185 Rome, Italy.\\ 
$^2$Laboratoire de Physique de la Mati\`ere Condens\'ee, C.N.R.S. Ecole Polytechnique, 91128 Palaiseau, France.\\ $^3$ Centre de Math\'ematiques et de Leurs Applications, Ecole Normale Sup\'erieure, 94235 Cachan, France.}

\begin{abstract}
We discuss various situations where the formation of rocky coast morphology can be attributed to the retro-action of the coast morphology itself on the erosive power of the sea. Destroying the weaker elements of the coast, erosion
can creates irregular seashores. In turn, the geometrical irregularity participates in the damping of sea-waves, decreasing their erosive power. There may then exist a mutual self-stabilization of the wave amplitude together with the irregular morphology of the coast. A simple model of this type of stabilization is discussed. 
The resulting coastline morphologies are diverse, depending mainly on the morphology/damping coupling. In the limit case of weak coupling, the process spontaneously builds fractal morphologies with a dimension close to $4/3$. This provides a direct connection between the coastal erosion problem and the theory of percolation. 
For strong coupling, rugged but non-fractal coasts may emerge during the erosion process, and we investigate a geometrical characterization in these cases.  The model is minimal, but can be extended to take into account heterogeneity in the rock lithology and various initial conditions. This allows to mimic coastline complexity, well beyond simple fractality.
Our results suggest that the irregular morphology of coastlines as well as the stochastic nature of erosion are deeply connected with the critical aspects of percolation phenomena. 

\end{abstract}


\end{frontmatter}

\section{Introduction}
\label{sec:introduction}

Coastline morphology is of current interest in geophysical research
and coastline erosion may have important economic
consequences~\cite{eric,penning}. Even more, the concern about global
warming has increased the demand for a better understanding of coastal
evolution.  This paper deals specifically with the erosion of rocky
coasts.

Rocky coasts have been estimated to represent $75\%$ of the world's
shorelines~\cite{davies}. They are found in different contexts, and
there exists a rich bibliography on the subject, see for
example~\cite{finkl} and the references therein, as well
as~\cite{naylor2010} for an update bibliography.  However, this
estimation strongly depends on the very definition of what constitutes
a rocky coast. Many cliffed coasts are fronted by beaches, with many
different morphologies and several different dynamical processes in
action. The morphology of these sea-shores may result from several
different processes; tectonicity and various erosion mechanisms (sea,
rivers, wind) acting on different soils as well as the possible role
of sediment transport. Sea erosion can be imperceptibly slow, but
nevertheless shapes coastal morphology. It can also be observed over
human timescale, being of concern to planners. For instance, a study
of cliff coasts in New Zealand reports sea erosion rates with peaks as
high as $9$ meters in one year~\cite{Gibb1978,Gibb1984,Kennedy2007}
(with typical rates ranging from $0.02$ to $0.5$ m/yr). Eroding
cliffed shorelines account for $20\%$ of the entire New Zealand
coast~\cite{Kennedy2007}.

Accordingly to the definition of rocky coasts given
in~\cite{naylor2010}, here we address coasts dynamics in the limiting
case where the role of sediment transport is considered to be
negligible. For instance, tectonically active coasts often display
rocky coasts with very limited sediment deposited by rivers (as in
Peru and Chile or along the North America cordillera). The rugged
appearance of these coasts is usually considered as an extension of
the rugged mountains characterizing the nearby landscape, however it
is difficult to exclude that sea wave erosion does not play a role in
their morphology. Collision coasts also tend to be rocky, containing
few depositional features. Because of their relative youth,
neo-trailing edge coasts, such as the Arabian coast along the Red Sea,
are also rugged and mostly rocky. Furthermore, there are many sites
throughout the world where rocky and rugged coasts are found in
tectonically passive margins, such as South Africa, parts of Argentina
and Brazil, eastern Canada, southern Australia and a section of
north-west Europe.

One might think that wave erosion can play a role in relatively low
rocky coasts, but the height of the cliff is not a general
contraindication for erosive sea dynamics. Very often these coasts
exhibit some kind of irregular morphology. In the last decades,
attempts to describe global geometry of sea-coasts have been made
using the tools of fractal geometry to the point that the coast of
Britain has been taken as an introductory archetype of self-similarity
in nature~\cite{mandelbrot67,mandelbrotbook,mandelbrot75}. Since then,
many tentative applications of fractal concepts to
geomorphology~\cite{geomorphologyvol91,Xu1993245} have been published but at the
same time there has been some debate about the fractality of
coastlines~\cite{goodchild,andrle,bartley}.

In other words, coasts and rocky coasts may be fractal or not,
depending possibly on the scale on which the coast is observed. Often
but not always, different scales exhibit different shapes. This
corresponds to the variety of possible contributions to coast
morphology mentioned above. Nevertheless, the observation of
geometrical similarities and the presence of "some sort" of scale
invariance in the morphology of rocky coasts, may suggest the
existence of a common mechanism which, when in action, shapes this
type of coastline.

A qualitative model for the appearance of fractal sea-coasts had been
suggested in~\cite{sapo1}. The idea is that irregular coasts
contribute to the damping of sea waves with the consequence that the
resulting erosion is weaker.  More recently, a numerical model of such
coastal dynamics was developed~\cite{sapoprl} and studied. It was
found that a mechanism based on the retro-action of coastal shape on
wave damping, leads necessarily, for the specific case worked in this
paper, to the self-stabilization of a fractal
coastline. Interestingly, it was found that the self-stabilized
geometry belongs to a well defined universality class, precisely
characterized in the mathematical theory of
percolation~\cite{stauffer}. In that sense, the notion of fractal
geometry for rocky sea-coasts should no more appear as a curiosity,
but as a necessary consequence of percolation theory.

In this paper we advocate the model, discussing the statistical
analysis of earth geomorphological data, the possible role of
sediments in erosion, the role of large scale heterogeneity in the
lithology of the eroded coasts, and we address the problem of
statistical characterization of different non fractal morphologies
predicted by the model. The structure of the paper is the following.

In Section~\ref{sec:singular} we present some statistical field
geomorphological data that suggest that coastlines geometry differs
from the general earth geomorphology.

Our specific erosion model, based on the aforementioned retro-action
mechanism, is presented in Section~\ref{sec:model}.  There we discuss
the fundamental connection between our model and the general theory of
percolation.

The erosion dynamics, short and long term evolutions, are discussed in
Section~\ref{sec:dynamics}. We also discuss how the critical nature of
percolation, results in a very irregular, or episodic, erosion
process.

The question of fractal versus non-fractal sea-coast is discussed in
Section~\ref{sec:fractalornot}. More precisely we give statistical
arguments that could help to distinguish transitory from final
morphology.

In Section~\ref{sec:geology} we address the fundamental question of
the role of geology in the frame of our retro-action model. The
qualitative results is that, very often, coastal morphology should be
the results of both retro-action and geological constraints.

In Section~\ref{sec:sediment} we discuss the possible role of
sediments and rubbles in the erosion process. We show that in various
cases the same scaling and morphologies should be obtained.

In Section~\ref{sec:complexity} we present some detailed data analysis
of some rocky coasts, unfolding the complexity of the real morphology
of coastlines as well as of the inland coastal regions.

In Section~\ref{sec:conclusion} we give the summary and conclusion of
the paper.

\section{ Is coastline morphology "distinct" in global geomorphology?}
\label{sec:singular}

A question that can be raised, observing rocky coasts, is whether
their geometry is simply inherited by the morphology of the
inland~\cite{mandelbrot75}, or whether the interaction with the sea
changes and shapes the coast in a distinct way.  In order to
disentangle and recognize specific features of coastlines, with
respect to higher earth surface isolines, it is interesting to use
tools which may help to reveal general universal features of their
geometry.

\subsection{Coastline morphology in global geomorphology}

\begin{figure}[h]
\centerline{\includegraphics[width=14cm]{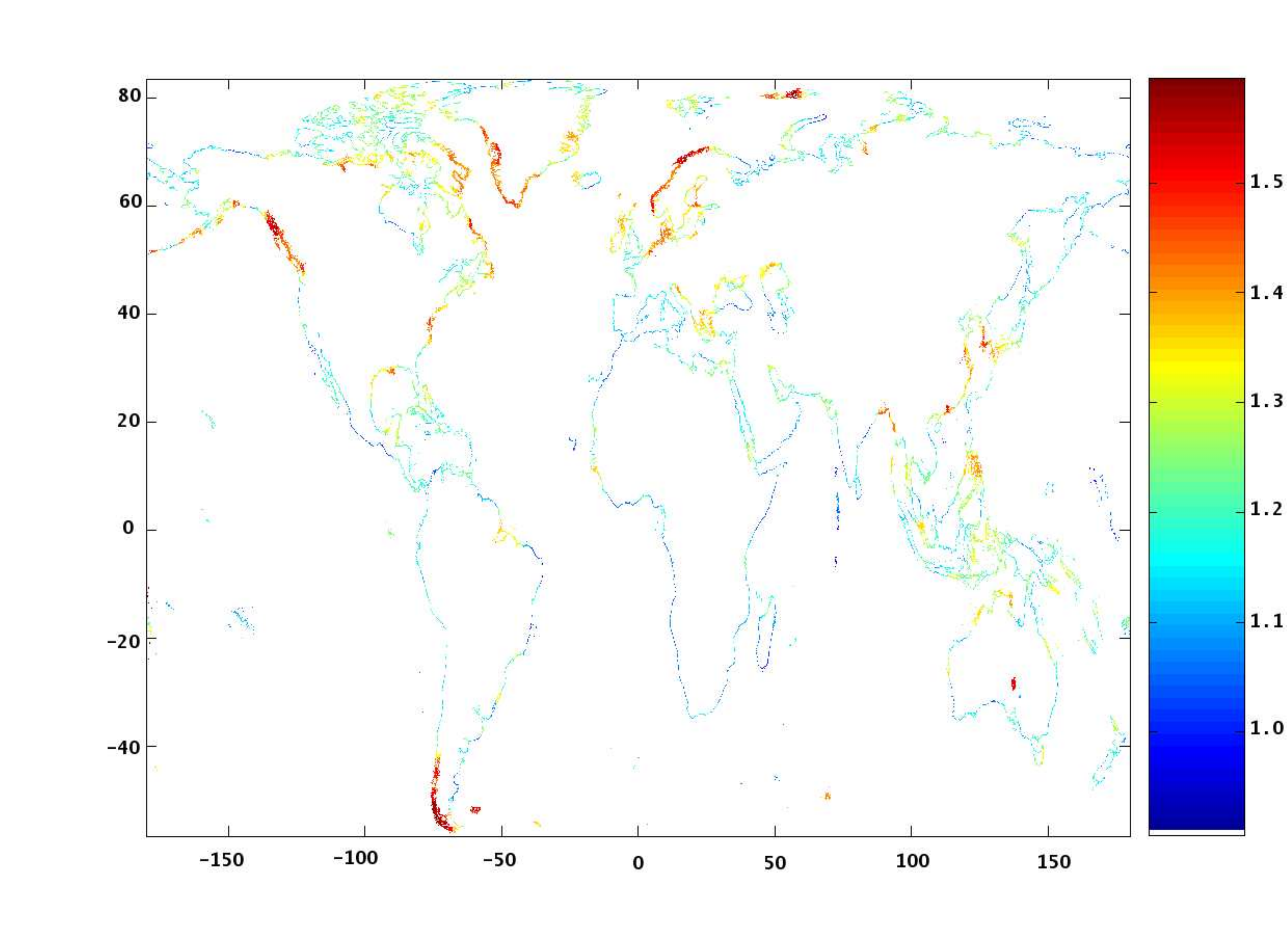}}
\centerline{\includegraphics[width=14.0cm]{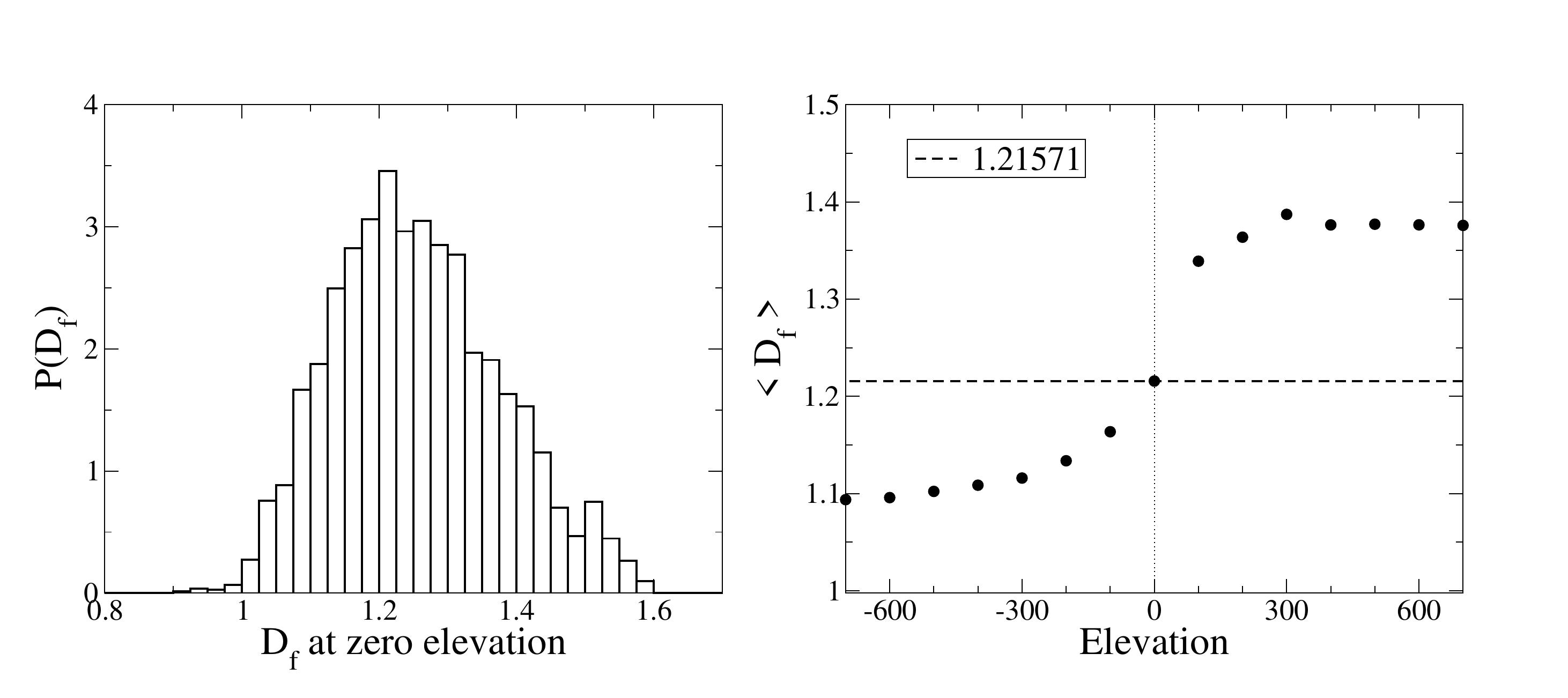}}
\caption{\label{worldiso} Statistical analysis of coastlines fractal
  dimension. Top: color map of earth coastline, the color shows the local measured fractal dimension. Bottom left: Distribution of measured fractal dimension on the Earth coastlines.  Bottom right: average fractal dimension for Earth  isolines as a function of elevation (in meters)
  between $-700m$ and $700m$; the world average fractal dimension for
  coastlines (elevation $0$) is slightly above 1.2.  The horizontal
  dash-dotted line corresponds to $4/3$. }
\end{figure}

The following analysis of the world coastlines is part of a more
general and detailed analysis, published elsewhere~\cite{fractalworld}
(more details in the Appendix~\ref{app:isolines}). Topographic data
for Earth have been obtained from the SRTM30-plus
set~\cite{srtm-plus}.  The data consists in the earth surface
elevation over a grid of points. From the data we computed earth
isolines, the coastline belonging to the $0$ elevation isoline.  Next,
the whole Earth surface is divided in squares of $4$ degrees latitude
x $4$ degrees longitude and the fractal dimension is computed for each
isoline portion in each square.  Fig.~\ref{worldiso} shows the result
of the coastline fractal analysis.  The Bottom left panel of
Fig.~\ref{worldiso} represents the distribution of the measured
fractal dimensions for the zero elevation isoline. The world coastline
average dimension is found to be slightly above $1.2$, but rocky
coasts have often higher dimensions.

Moreover, one can consider the behavior of the global isolines as a
function of elevation near the sea level and compare their
corresponding dimension. In the bottom right panel of
Fig.~\ref{worldiso} we show the average fractal dimension measured
between $-700$ and $700$ meters. Interestingly, exactly at $0$
elevation a rapid change in the measured average fractal dimension is
observed. This gives an indication that the interaction between sea
and land is responsible for the complex geometry of coastlines. In
this sense, a model for the geometry of rocky coastlines should
explicitly take in account the main physical processes taking places
at the interface between sea and land, that is the dynamics of coastal
erosion.

\subsection{The specific case of plateau coasts}
\label{sec:plateau}

The global analysis presented so far, suggests that the coastal
morphology is not the sole reflect of the inland morphology. Large
scale geometry of coastline may be the result of many different
characteristic phenomena. Sand deposits usually smooth the
irregularity of rocky coasts, filling bays, or may display specific
patterns~\cite{murraynature}. On the other hand the rough geometry of
glacial valleys gives the very convoluted coastline typical of fjords
at large absolute latitudes.

\begin{figure}[ht]
\centerline{\includegraphics[height=5cm]{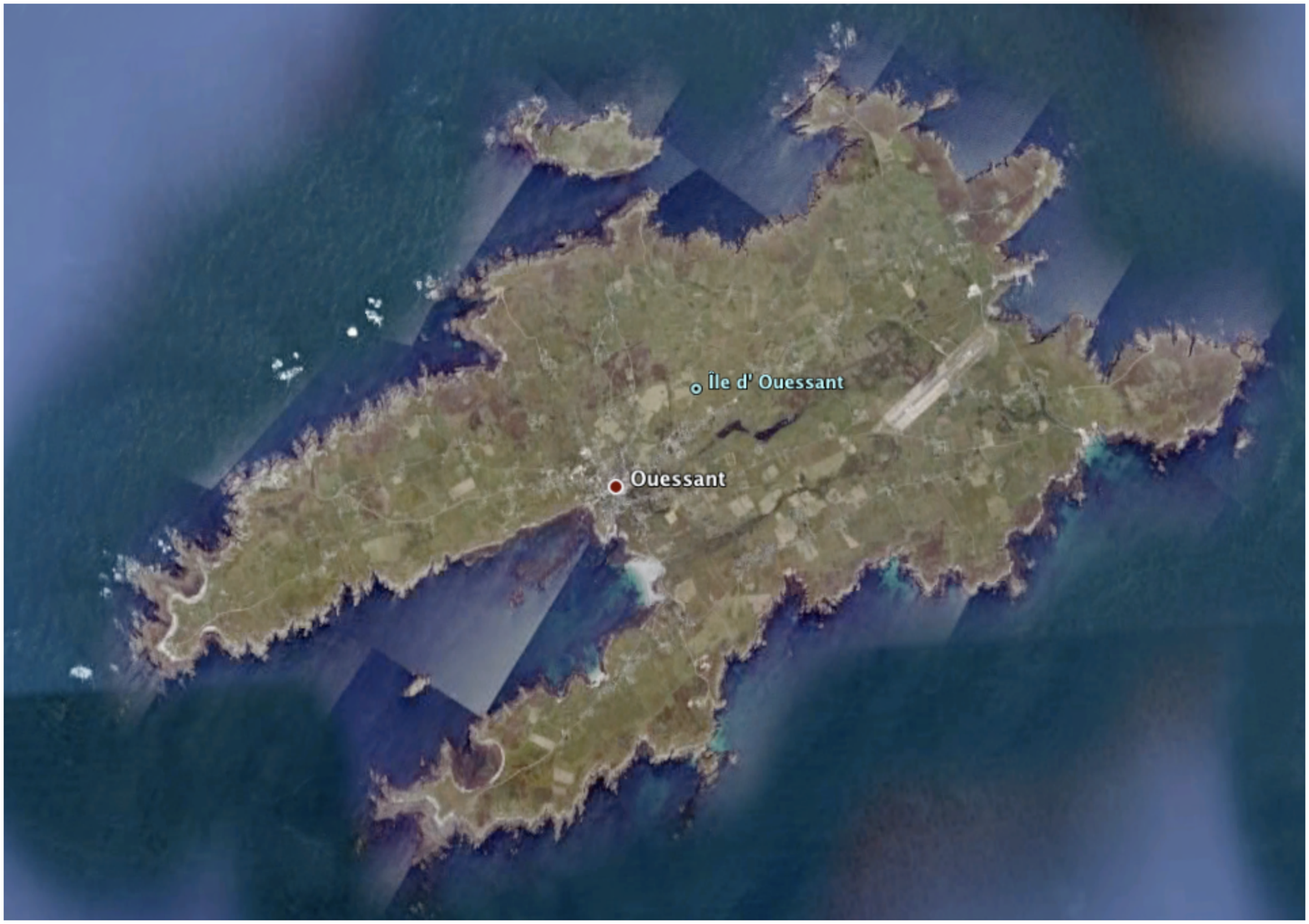}
\includegraphics[height=5cm]{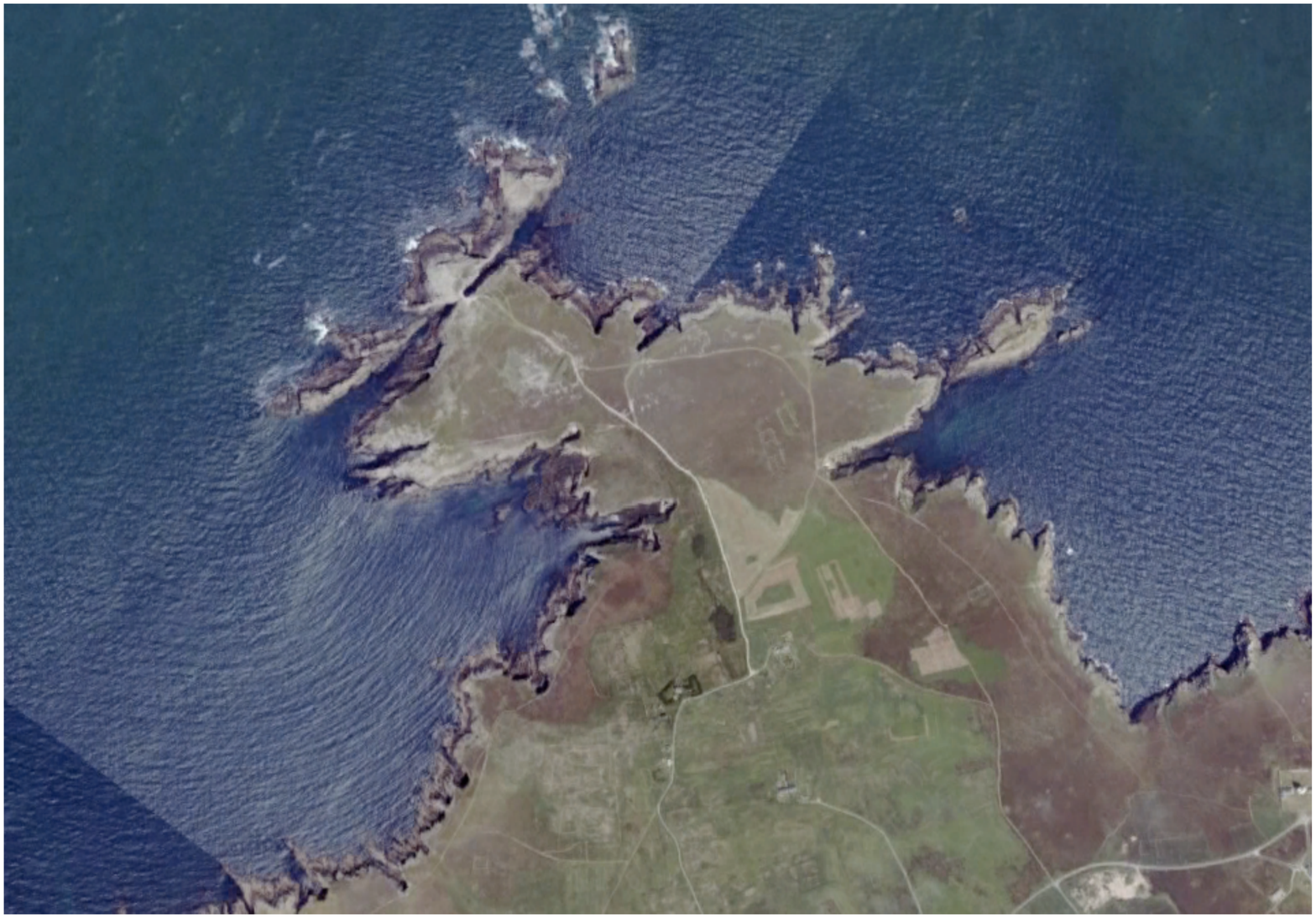}
}
\centerline{
\includegraphics[height=5.5cm]{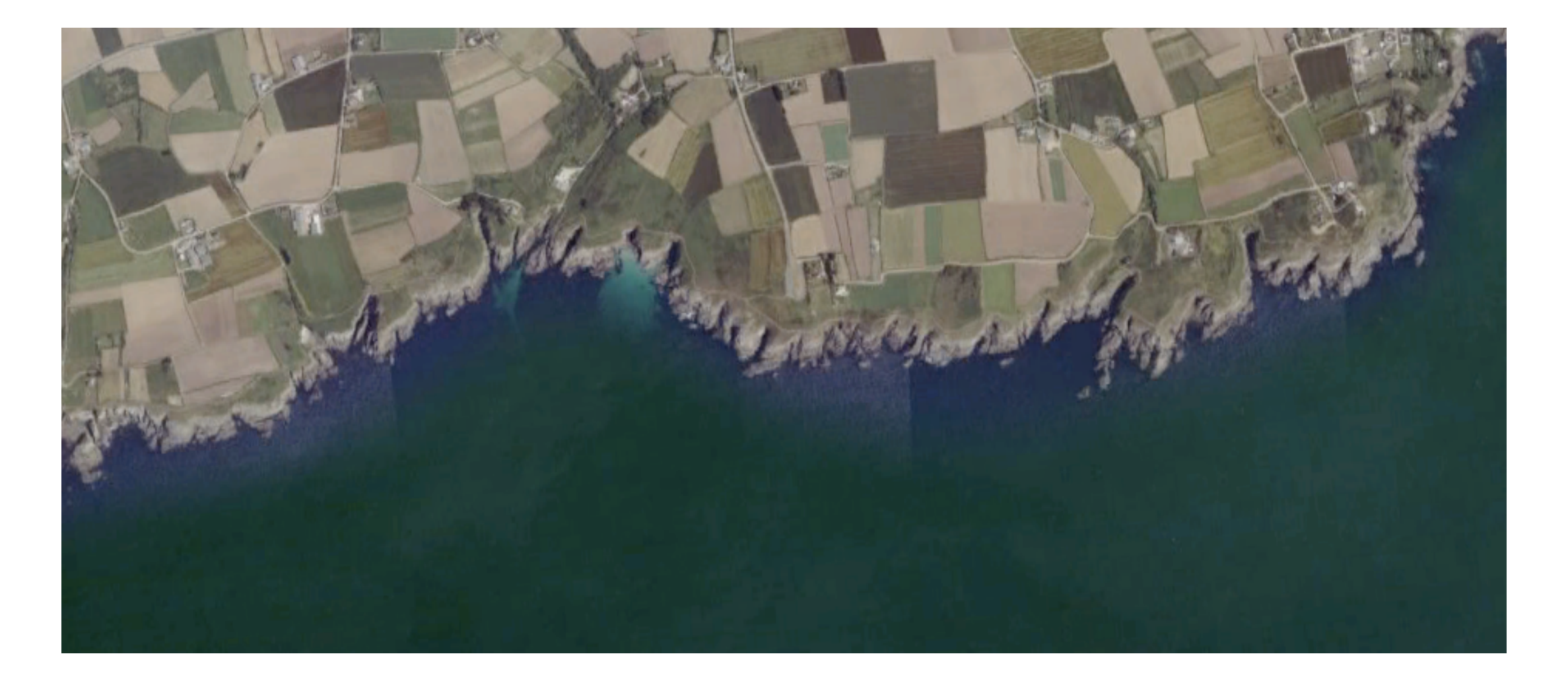}
}
\caption{\label{fig:plateau} Example of plateau coasts. Top left: Ouessant Island, Brittany, France; top right: detail of the north east coast of the island. Bottom: coast at the south of Plougonvelin, Brittany, France. Note the cultivated fields right to the sea-shore. (Pictures are snapshot from Google Earth)}
\end{figure}

However, in many cases, rocky coasts present a very {\em steep slope}
(or terrain gradient) with respect to the inland profile. This is the
case of cliffs, or what we call "plateau" coasts. In our mind, a
plateau coast is characterized by an extreme situation where a flat
landscape becomes steep at the coast. A photographic example is shown
in Fig.~\ref{fig:plateau}.  In this case one expects: first, that sea
erosion is the most relevant shaping mechanism, and second, that a 2D
model (as the one presented in the next section) could be sufficient
to approach such morphology evolution.

Of course, the case of plateau coasts is a limiting case. Discussion
of several other examples of coastline complexity will be given in
Section~\ref{sec:complexity} below. Through these more detailed
examples, we wish to put forward the idea that those shores with {\em
  high local slopes (or terrain gradients) at the shore} may be
reasonably attributed to marine erosion, acting in a given geological
context.


\section{The 2D model of rocky coasts erosion}
\label{sec:model}

Rocky coasts erosion is the product of marine and atmospheric
causes~\cite{davies2}. There exist many different erosion processes:
wave quarrying, abrasion, wetting and drying, frost shattering,
thermal expansion, salt water corrosion, carbonation, hydrolysis. In
the same time, the mechano-chemical properties of the rocks
constituting the coast, which are linked to structure, composition and
aging defining their ``lithology'', exhibit an unknown dispersion.

On the other hand, erosion is the consequence of the existence of an
``erosion power''.  It is a selective mechanism, which progressively
eliminates the weaker parts of the surface. The remaining shore is
then hardened as compare to the initial shore. But the erosion power
is not constant and may change during erosion. In particular, damping
mechanisms caused by the erosion itself could arise, establishing a
self-stabilizing mechanism. In other words, erosion is the product of
three ingredients: erosion power, rocks lithology, {\em and} a damping
depending on the shore morphology. The interplay between these three
factors is discussed here in various conditions, through the numerical
implementation of a simple model~\cite{sapoprl}.

The specific damping mechanism considered here, relies on the studies
of irregular or fractal acoustic cavities. They show that viscous
damping is increased on a longer, irregular
surface~\cite{sapo2,sapo3,simon,simon2}. These considerations have
been applied practically in the conception of efficient acoustic road
absorber now installed along several roads in France~\cite{patent}.

An other example of the proposed selection mechanism, is the case of
the dynamics of pit corrosion of thin aluminum
films~\cite{balasz}. There, fractal geometries spontaneously appears
at the interface of the corroded solid. The phenomenon can be
understood by means of a minimal model~\cite{sapo5}, which disregards
many atomic details of the corrosion process. The analogy between
erosion and corrosion is less artificial than one could think {\em a
  priori} (see discussion in Section~\ref{sec:sediment}).

In the following we make an arbitrary distinction between {\it
  ``rapid''} mechanical erosion (namely wave quarrying) and {\it
  ``slow''} weakening of the rocks due to the action of the elements
(weathering processes). These {\it ``slow''} weakening events trigger,
from time to time, new {\it ``rapid''} erosion sequences. The
justification is that mechanical erosion generally occurs rapidly,
mainly during storms, after rocks has been slowly altered and
weakened. We first study this supposedly rapid erosion
mechanisms. Then we show that the full complex dynamics, involving
fast and slow processes, changes the shape of the coast on a longer
time scale keeping its gross geometrical characteristics. This
dynamics is reminiscent of the {\em quasi-equilibrium} evoked by
Trenhaile~\cite{Trenhaile2002} (see below)

Our model schematize the sea, the land, and their interaction in the
following way.

\subsection{The sea-coast system as a damped resonator}
\label{sec:resonator}

In analogy with the acoustic oscillations in a cavity, the sea,
together with the coast, is considered to constitute a
resonator~\cite{sapo2,sapo3}. It is assumed that there exists an
average excitation power of the waves $P_0$. The ``force'' acting on
the unitary length of the coast is measured by the square of the wave
amplitude $\Psi^2$.  This wave amplitude is related to $P_0$ by a
relation of the type $\Psi^2\sim P_0 Q$ where $Q$ is the morphology
dependent quality factor of the system: the smaller the quality
factor, the stronger the damping of the sea-waves. There are several
different causes for damping. Since the different loss mechanisms
occur independently, the quality factor satisfies a relation of the
type
\begin{equation}
\frac{1}{Q}=\frac{1}{Q_{coast}}+\frac{1}{Q_{other}}\,,
\label{eq1}
\end{equation}
where $Q_{coast}$ is the quality factor due to the ``viscous''
dissipation of the fluid moving along the coast and the nearby islands
and $Q_{other}$ is related to other damping mechanisms (e.g. bulk
viscous damping).  Studies of fractal or irregular acoustic
cavities~\cite{sapo2,sapo3,simon,simon2} have shown that the viscous
damping increases roughly proportionally to the cavity perimeter.

\begin{figure}[h]
\centerline{\includegraphics[height=4.2cm]{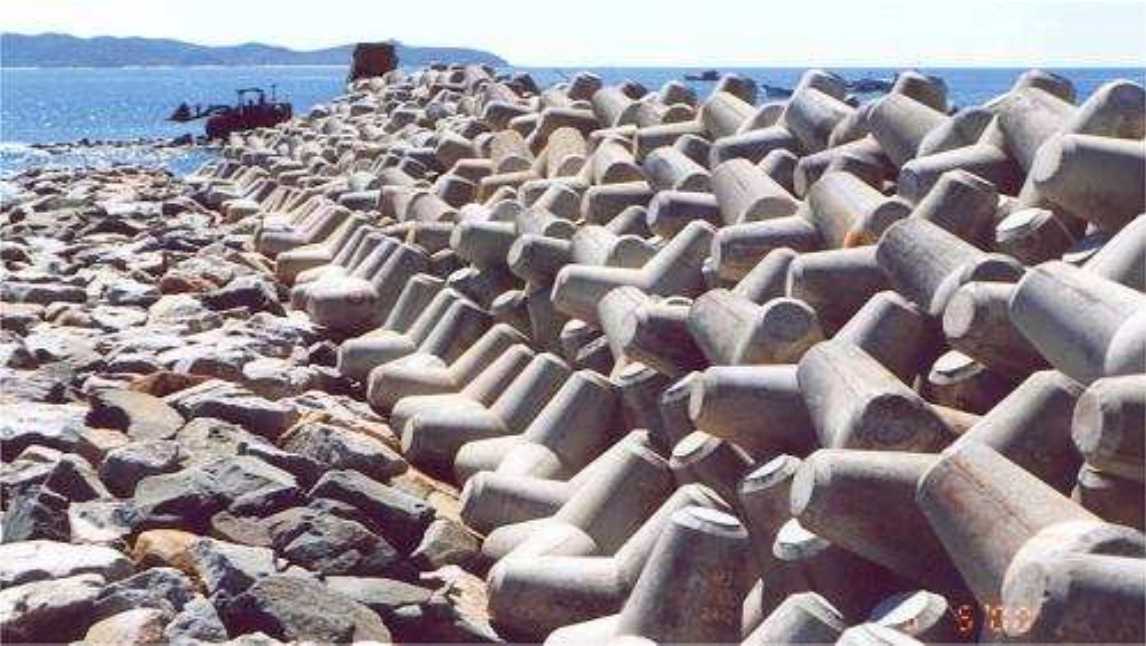}\includegraphics[height=4.2cm]{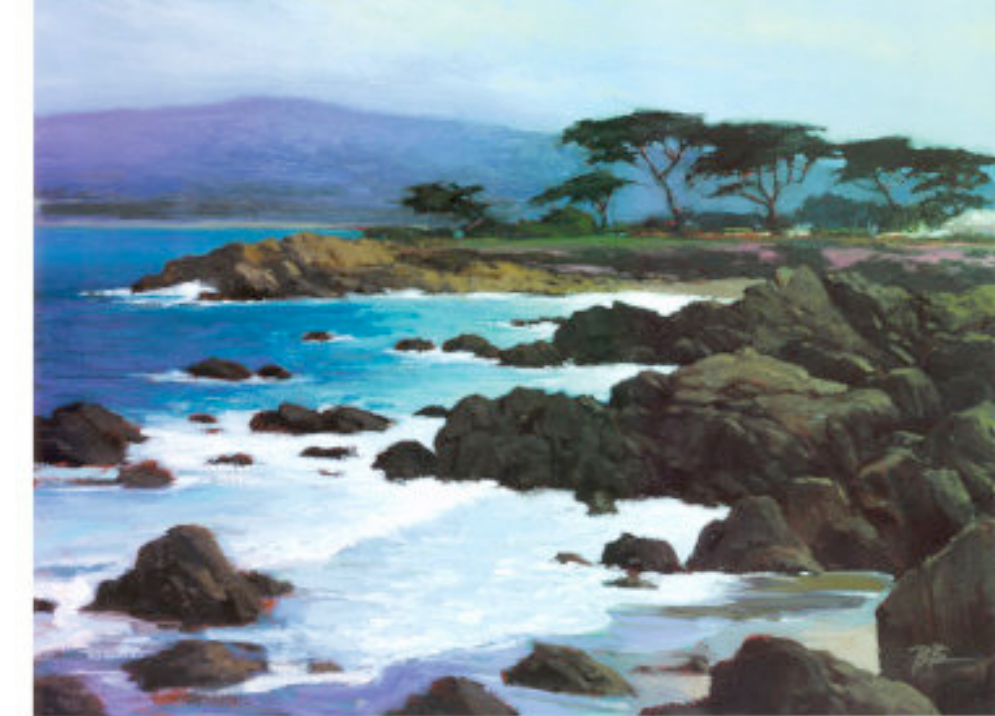}}
\caption{\label{fig:tetrapods} Artificial and natural break-waters. Left: Break-water made of concrete tetrapods on
layers of rocks (picture from http://www.sys.com.my). Right: Picture taken at Pacific Grove, in the Monterey peninsula, California. Our measure of the fractal dimension of such coast is close to $4/3$.}
\end{figure}

This model uses, as a working hypothesis, the idea that sea-waves are
more damped along an irregular coast much in the same way as acoustic
modes in irregular cavities. This appears to be an empirically known
effect used to build efficient break-waters that are based on
hierarchical accumulations of tetrapods piled over layers of smaller
and smaller rocks, in close analogy with fractal geometry (see
Fig. \ref{fig:tetrapods}, left, and the many descriptions of
breakwaters in Ref.~\cite{manual}).

\begin{figure}[h]
\centerline{\includegraphics{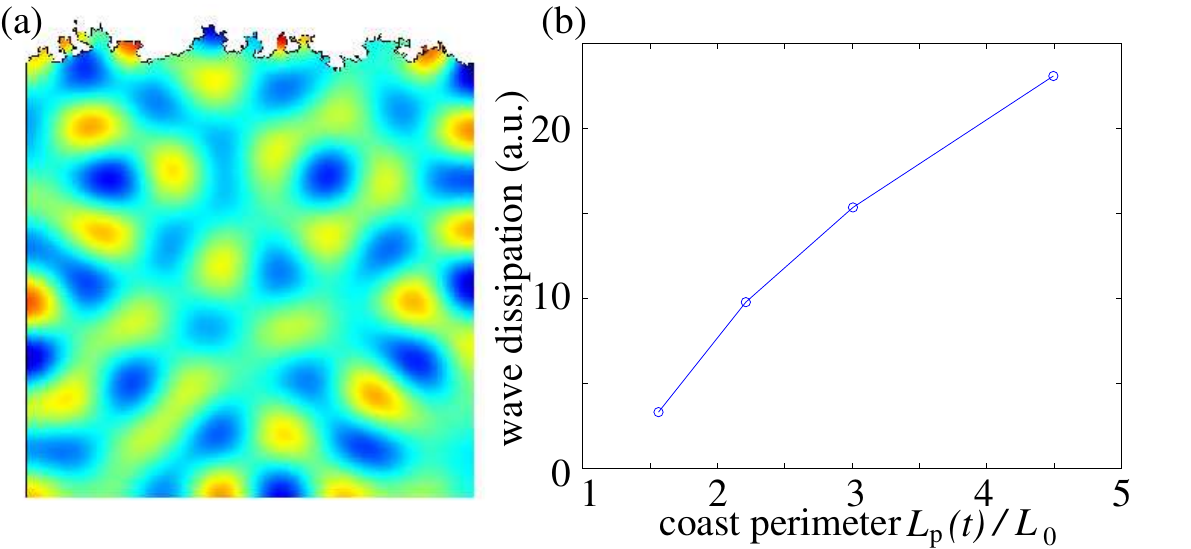}}
\caption{\label{fig:cavities} (a) A numerically computed eigenmode in
a 2D resonator with the upper side representing one of the coast
shapes during erosion. The energy dissipation of a mode is proportional to
the integral of the squared amplitude along the
coastline~\cite{simon,simon2}. Frame (b)
shows the evolution of the average dissipation for four different
resonators, whose upper boundary corresponds to four successive times.
During erosion, the coast perimeter $L_p(t)/L_{0}$
increases and the damping increases roughly proportionally to the
coast perimeter.}
\end{figure}

To check the relation between irregularity and damping for the
particular shapes of a sea-coast during erosion, we consider the
properties of four 2D resonators, with the upper side corresponding to
four successive coast shapes.  For each time and morphology we solve
numerically the Helmoltz wave equation with Neumann boundary
conditions.  We compute the eigenmodes in a given frequency range,
assuming weak losses on the eroding profile.  For each eigenmode we
compute the energy dissipation which is supposed to take place on the
coast. In other words we study the wave dissipation due to viscous
forces acting on the boundary of irregular swimming pools.  Results
are given in Fig.~\ref{fig:cavities}, showing that the average losses
of the computed eigenmodes in the four resonators increase roughly
proportionally to the coast perimeter. The fact that the velocity of
sea waves depends on the sea-floor depths would not modify the general
link between perimeter and damping.

Therefore, one can, in first approximation, assume that $Q_{coast}$ is
inversely proportional to the coast perimeter $L_p(t)$ whereas
$Q_{other}$ is independent of the coast morphology.  In other words,
the sea exerts a homogeneous erosion power $f(t)$ on each coast
element proportional to $\Psi^2(t)$:
\begin{equation}
f(t)=\frac{f_0}{1+\frac{g\,L_p(t)}{L_0}}\,,
\label{eq2}
\end{equation}
where $L_p(t)$ is the total length of the coast at time $t$ (then
$L_p(t=0) =L_0$). The factor $g$ measures the relative contribution to
damping of a flat shore as compared with the total damping. The
quantity $f_0$ is the renormalized value of $P_0$ such that $f(t)<1$
at all $t$. Small $g$ factor correspond to {\em weak coupling} between
the erosion power and the coast length and large $g$ correspond to
{\em strong coupling}.

Note that for the higher frequency waves with short wavelengths that
contains a large erosion power, it is clear that their damping is
proportional to the coast perimeter. These are the breaking waves
usually considered to be the most erosive waves.

The functional dependence of the erosion force as a function of the
coast perimeter could be different without affecting the
results. Rather, a better model for damping should take care of a
possible wave frequency dependence as well as the possibility of
localization effects along the irregular coast~\cite{sapo2}.  This
would modify Eq.~\ref{eq2} and change the time evolution. Note
however, that what is important here is that, as erosion proceeds and
sea "penetrates" progressively the earth, the erosion power is
diminished.  This is the essence of such a retro-action model. Any
model which would present this property would lead to the same type of
results. In particular if erosion sediments stay locally on the sea
floor, (are not transported) and contribute to damping, the erosion
power would decrease as a function of the total amount of material
already eroded and would create the same type of effects (see below).

\subsection{The land as a disordered solid}
\label{sec:land}
\begin{figure}[h]
\centerline{\includegraphics[width=8.5cm]{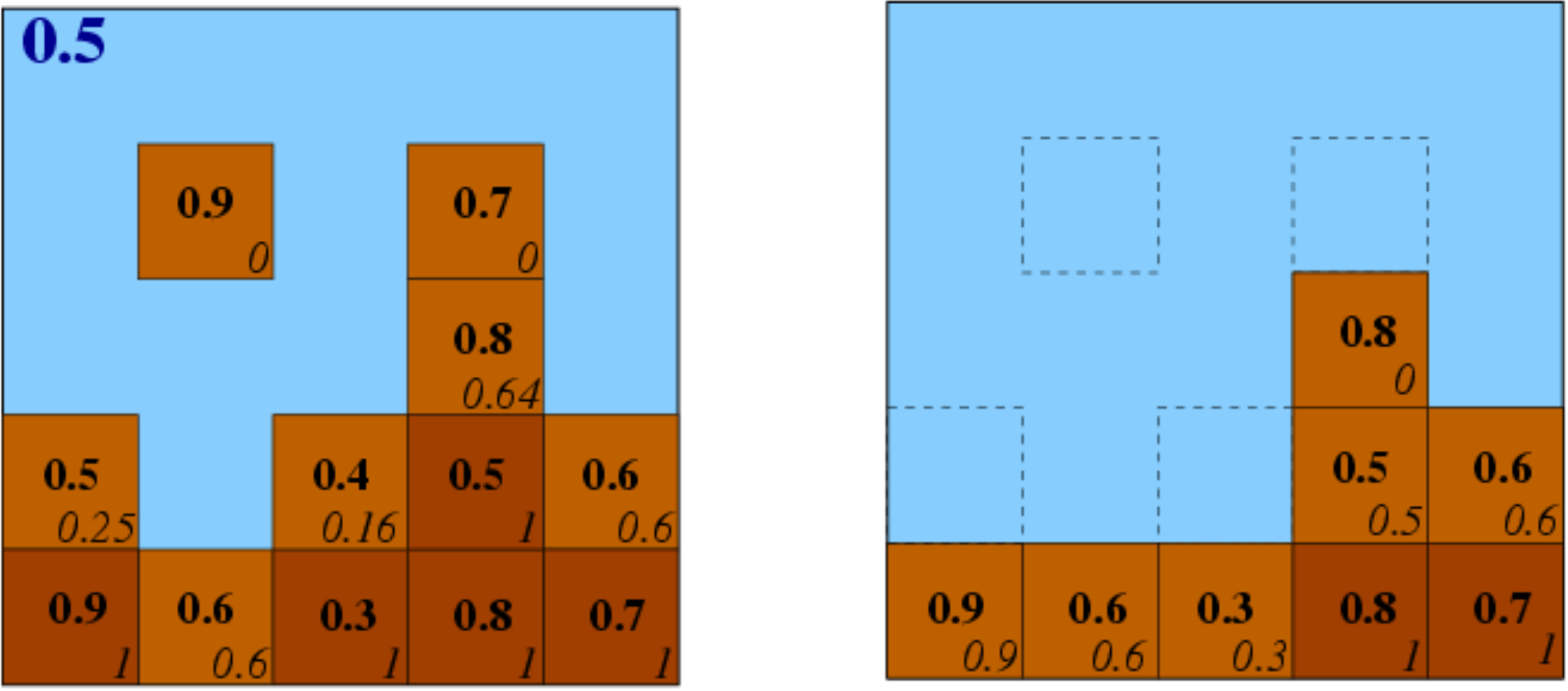}}
\caption{\label{fig2} Illustration of the erosion process.  The thick
number at the square center represent the lithology $\{l_i\}$.  The
numbers in the corners are the corresponding resistances $r_i$ which
depend on the local environment as explained in the text. The sites
marked with 1 are earth sites with no contact with the sea.  Left and
right: situations before and after an erosion step with
$f(t)=0.5$. After this step resistances are updated due to the new sea
environment.}
\end{figure}

The ``resisting'' random earth is modeled by a square lattice of
random units of global width $L_0$. Each site represents a small
portion of the earth, named {\em a rock} here. The sea acts on a
shoreline constituted of these rocks, each one characterized by a
random number $l_i$, between $0$ and $1$, representing its
lithology. The erosion model should also take into account that a site
surrounded by the sea is weaker than a site surrounded by earth
sites. Hence, the resistance to erosion $r_i$ of a site depends on
both its lithology and the number of sides exposed to the action of
the sea. This is implemented here through the following weakening
rule: sites surrounded by three earth sites have a resistance $r_i =
l_i$. If in contact with $2$ sea sites the resistance is assumed to be
equal to $r_i = l_i^2$.  And, if site $i$ is attacked by $3$ or $4$
sides, it has zero resistance. The iterative evolution rule is simple:
at computer time step $t$, all coast sites with $r_i < f(t)$ are
eroded (sometimes exposing new sites to erosion), and then $L_p(t)$
and $f(t)$ are updated together with the resistances of the earth
sites in contact with the sea. Then, from one step to the next, some
sites are eroded because they present a ``weak lithology'' while some
strong sites are eroded due to their weaker stability due to sea
neighboring.  An example of local evolution is shown in
Fig.~\ref{fig2}. {\em Note that our variable $t$ simply denotes a
  number of computer steps and therefore is not a real time}.

\subsection{Results}
\label{sec:results}

\begin{figure}[ht]
\centerline{\includegraphics[width=14.0cm]{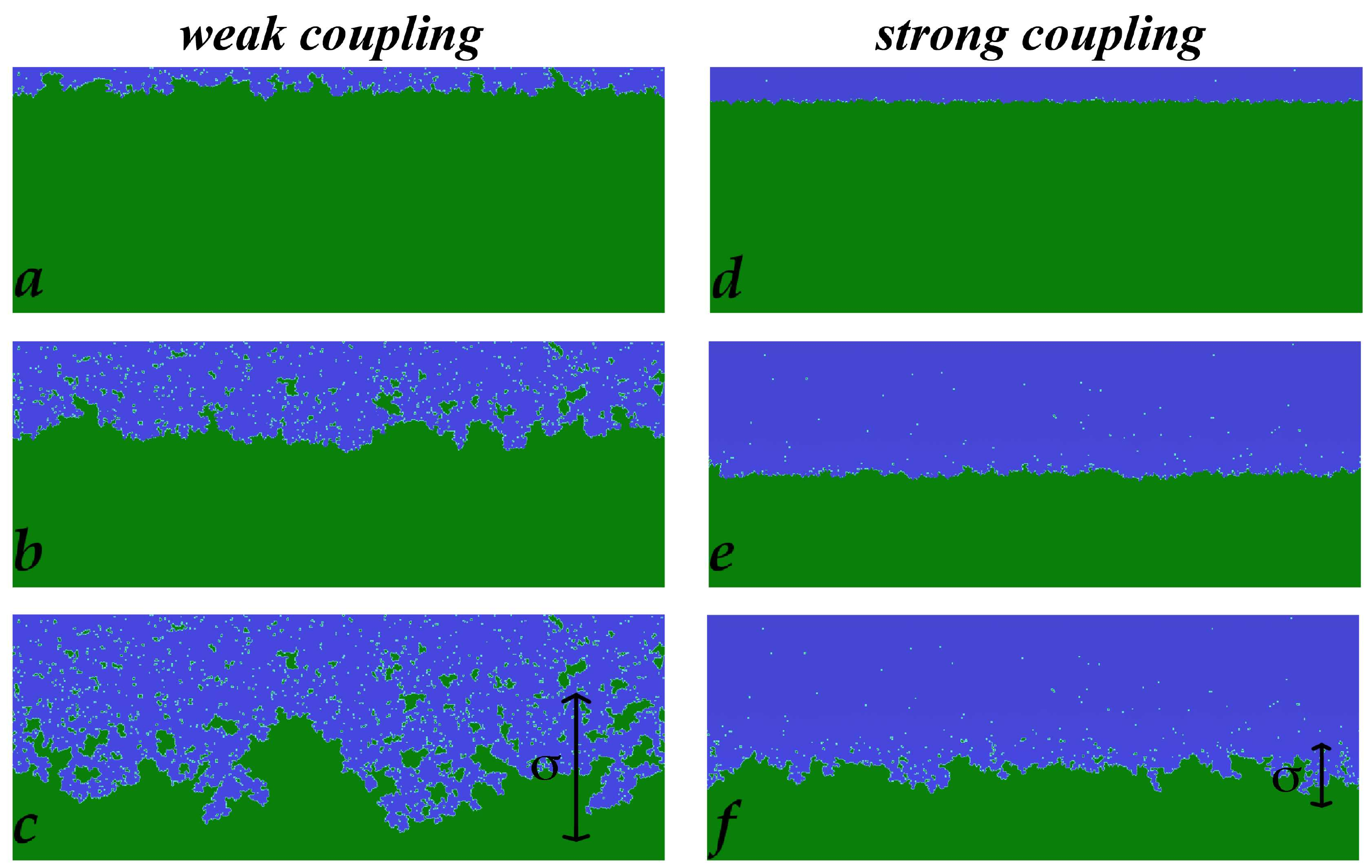}}
\caption{\label{fig1} Time evolution of the coastline morphology
starting with a flat sea-shore. Left and right columns respectively weak and
strong coupling. Top to bottom: successive
morphologies with the final morphologies at the bottom.
Note that case (b), transitory shape with weak coupling and case (f), final
shape with strong damping appear to be similar but it is shown below
that there exist statistical means to distinguish one from the other.}
\end{figure}

To exemplify the intrinsic properties of the model we consider an
artificial situation where erosion would start on a flat
sea-shore. The computer implementation of the above dynamic model
leads to a spontaneous evolution of the smooth seashore towards
geometrical irregularity as shown in Fig.~\ref{fig1}. The figure
exhibits the time evolution of an initially flat coastline towards
geometric irregularity. The left column describes the case of weak
coupling, the right column the case of strong coupling.

\begin{figure}[ht]
\centerline{\includegraphics[height=9cm]{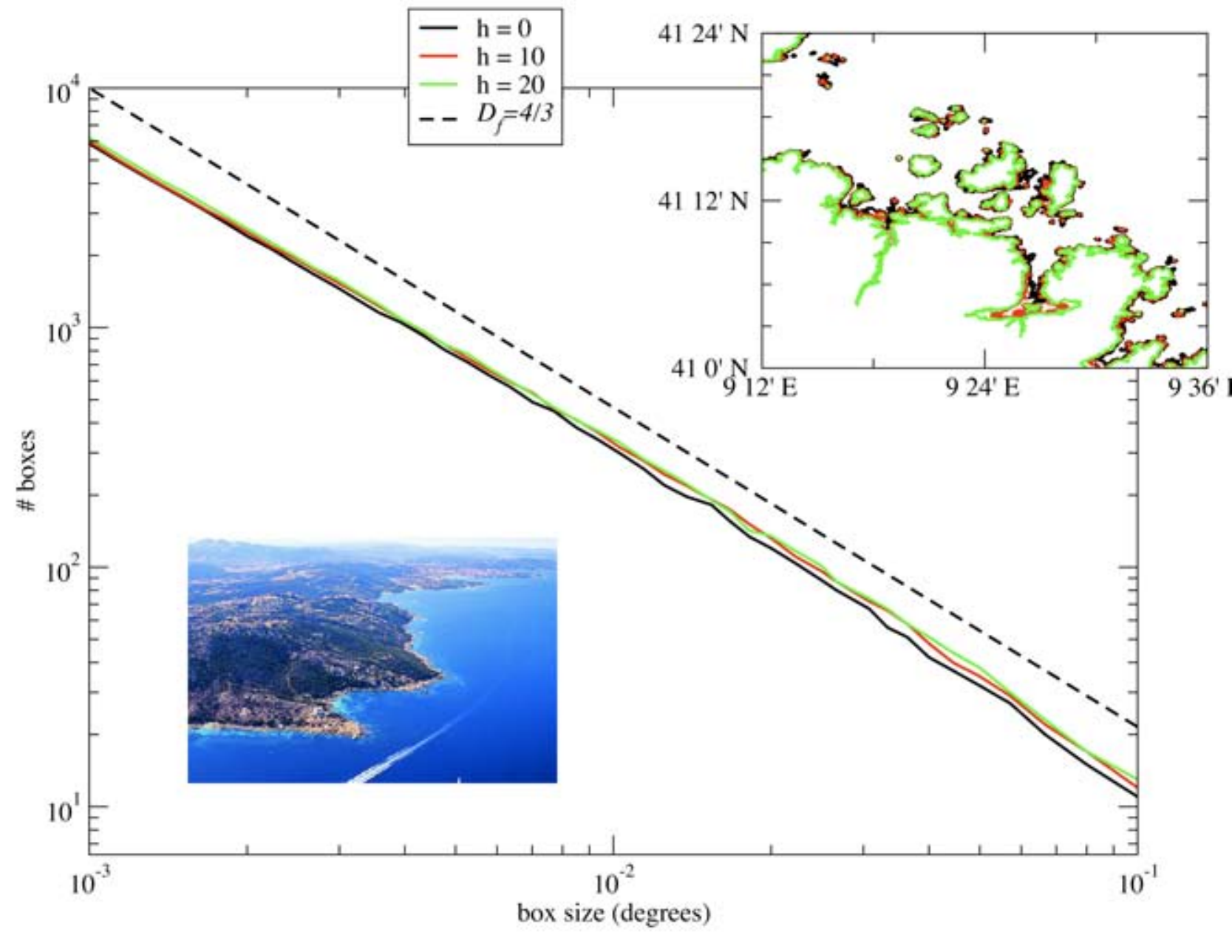}
}
\caption{\label{fig:palau} Northern coast of Sardinia (near Palau).  Top right: Box-counting measure of coastal isolines, compared with the ideal box-counting for a fractal with $4/3$ dimension. Top right inset: Coastal isolines, several elevations. Bottom left inset: A picture of the coast. \label{fig:sardinia}}
\end{figure}

In the case of {\em "weak coupling"} the terminal morphology is highly
irregular and it looks much like some of the irregular morphologies
observed on the field. Consider for instance, the north eastern coast
of Sardinia: the fractal dimension of this coast found to be very
close to $4/3$, as shown in Fig.~\ref{fig:palau}. There, we compare
the coastline fractal dimension ($0m$ isoline) and the isolines
closest to the coast, which is quite steep. On the opposite, the
inland does not present very high reliefs. The picture at the
bottom-left panel shows the coast near Palau.

\begin{figure}[ht]
\centerline{\includegraphics[width=10cm]{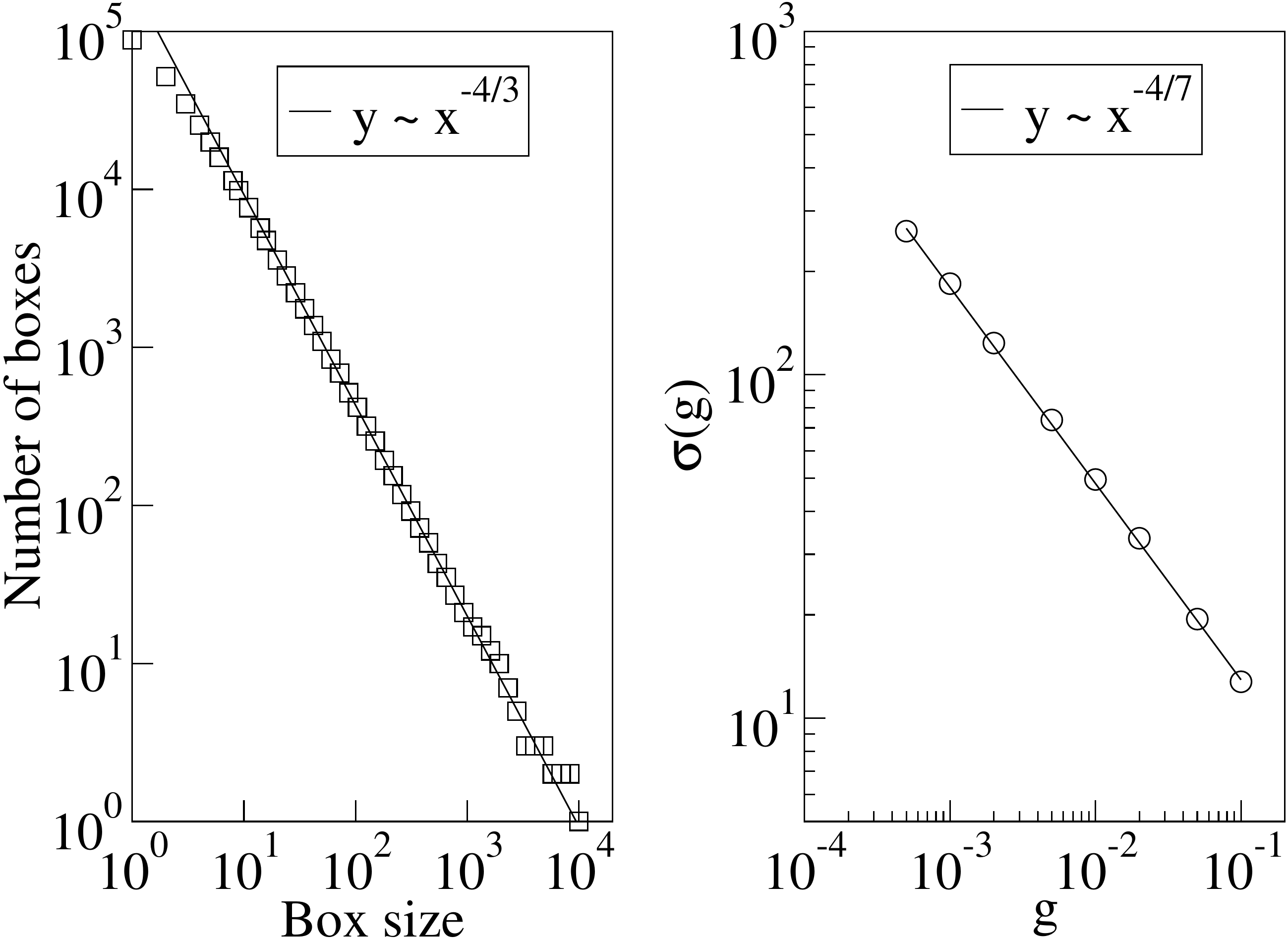}}
\centerline{\includegraphics[width=14cm]{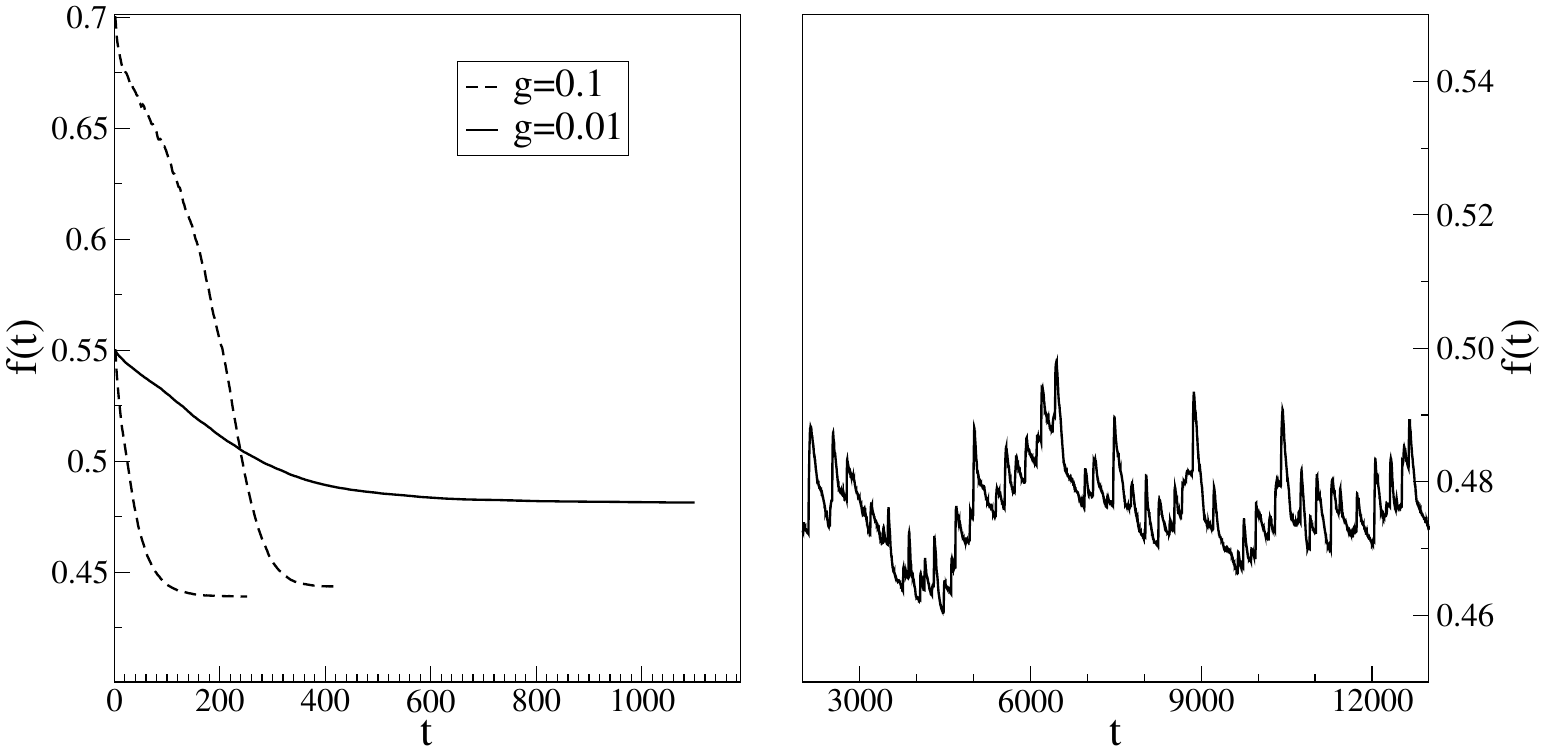}}
\caption{\label{fig4} Top left: Box-counting determination of the model coast
fractal dimension. The straight line is a power law with a slope
-4/3. The best fit gives: $D_f = 1.332(3)$. The data refer to a large
system with $L_0 = 10^4$, and small damping $g =10^{-4}$. Top right:
Scaling behaviour of the coast width $\sigma$. The straight-line is a
power law with the GP exponent -4/7 (each point is an average over
$400$ samples with $L_0 = 5000$). \label{fig3} Bottom: Dependence of the erosion force $f(t)$ as a function of the number of computer time steps $t$. The left figure shows the evolution of the sea erosion force acting on
the coastline during a ``rapid'' sea-erosion process (different
values of the scale gradients $g$ and of $f(0)$). This dynamics
spontaneously stops at a value weakly dependent on $g$ (systems with
$L_0 = 1000$, averaged over ten different realisations).  The right
plot is the erosion force $f(t)$ during the complete dynamics
(``slow'' weathering process triggering ``rapid'' erosion) illustrated
in Fig.~\ref{slow}.}
\end{figure}

Indeed, in the weak coupling case, our model produces a fractal
terminal morphology with a dimension very close to $4/3$ (see
Fig.~\ref{fig4}, left). Note, however, that the fractal morphology
extends up to a maximum scale, of the order of the transverse width of
the artificial coastline $\sigma$ (depicted in the last snapshots of
Fig.~\ref{fig1}). The precise mathematical definition of this
statistical width $\sigma$ is given below (see Eq.~\ref{sigmadef}).

Here we want to stress that the average width is the distance below
which the coast geometry is fractal. In other words we expect to have
a geometrical scale invariance up to a distance $\sigma$. At larger
scale the seashore can be considered as the reunion of independent
(uncorrelated) fractals of size $\sigma$.

It turns out that $\sigma$ depends directly through a power law of the
coupling parameter $g$. This is shown also in the right panel of
Fig.~\ref{fig4}.  One observes a power law with exponent close to
$-4/7$. As we will discuss below, the fractal dimension $D_f=4/3$ and
the $\sigma$ scaling exponent $-4/7$ are the signature of the deep
connection of our model with percolation theory. In the language of
statistical physics, the model belongs to the universality class of
percolation, as will be discussed in the next section.

For {\em ``strong coupling''}, the erosion ends on rugged
morphologies, as shown in Fig.~\ref{fig1}(f). This can be understood
as the limit of small $\sigma$, i.e. where the geometrical correlation
of the coast is short ranged. Note that such rugged morphology
resembles to transitory morphologies observed with weak coupling as
that of Fig.~\ref{fig1}(b).  We will see below in
Section~\ref{sec:fractalornot} that there exists statistical methods
to distinguish transitory (young coasts) from final morphologies (old
coasts).

In summary, we discuss a model which although based on very few
ingredients, gives rise to a variety of coastline morphologies,
fractal or simply rugged, transitory or final.


\subsection{Relation between coastal morphology and percolation theory}
\label{sec:percolation}

Our study indicates the existence of a connection between coastal
erosion of rocky coasts and percolation theory. Percolation is a
cornerstone of the theory of disordered systems, which has brought new
understanding and techniques to a broad range of topics in physics,
materials science, complex networks, epidemiology,
etc.~\cite{stauffer}

Percolation theory deals with the following statistical
problem. Consider a lattice, like the square lattice for instance, and
independently assign to each site a random number obtained from a
uniform distribution between $0$ and $1$. Now select the set of sites
which happen to have a number smaller than some fixed arbitrary value
$p$ and occupy them. If these selected sites are first nearest
neighbors, they define so-called clusters.  Of course if $p$ is small,
the clusters themselves will be of small size.  However, strictly
above a critical value $p_c$, there exists an infinite cluster that
crosses the lattice from left to right and from top to bottom.  The
most important characteristics of percolation phenomena is that near
criticality, that is when $p$ is close to $p_c$, the scaling
properties of the system are independent of the lattice geometry. Although the value of the percolation threshold $p_c$ does depend on
the lattice under consideration the exponents of the so-called scaling
laws that describe the properties of clusters and the geometry of the
infinite cluster at percolation are independent of the lattice. In
particular the external frontier of the percolation cluster is fractal
with a fractal dimension exactly equal to $7/4$ and the so-called
accessible perimeter has a fractal dimension equal to
$4/3$~\cite{duplantier,lawler,schramm,grossman}.

As the reader can suspect, there is a similarity between the
percolation cluster and the coastline produced by our model. In fact,
at the end of the erosion process, all earth sites at the interface
with the sea have a resistance larger than the wave erosive power. So,
in some sense, our erosion model spontaneously identifies, and stops
at, a percolating interface constituted of ``strong'' sites.

Indeed, there exist a direct relation between our model and the theory
of percolation. In particular, besides the fractal dimension of the
coast, the behavior of the width $\sigma$ as a function of the
coupling factor $g$ exhibit a power law dependance, as shown in
Fig.~\ref{fig4}. This power law is characteristic of a specific
variant of percolation, called {\em gradient
  percolation}~\cite{sapo4,rosso1,sapoval-intro}. In the
Appendix~\ref{app:gp} we explain gradient percolation and how it can
be related to our erosion model.

At this stage it is important to recall briefly the concept of
universality in statistical physics. Universality means that several
very different phenomena can exhibit analogous macroscopic properties
described by the same power law exponents. The phenomena which
exhibits the same exponents belong to a unique {\em universality
  classes}. The models described in this paper, percolation, gradient
percolation, our model of coastal erosion, all belongs to the
percolation universality class, irrespectively of many details. For
example, if we change the lattice geometry, or the distribution of the
rocks lithology, the model still belongs to the percolation
universality class.  Here it means that shorelines made of rocks of
different nature and sizes, subjected to different external climate,
can exhibit the same large scale geometrical properties.

Recent studies inspired by our model, corroborates the deep connection
between coastal morphology and percolation. A remarkable property of
such real coasts has been observed: they have been found to be
conformally invariant~\cite{boffetta}. This geometrical property
stands for itself but there exists a mathematical demonstration that
it exists for the so-called accessible perimeter of the percolation
cluster. (It should be stressed that not every fractal with dimension
$4/3$ is conformally invariant.~\cite{duplantier,lawler,schramm}).


\section{Dynamical evolution of the coast morphology}
\label{sec:dynamics}

Let now describe in more detail the erosion dynamics resulting from
the minimal rules defined by our model. For the sake of simplicity, we
consider first the case of a flat (smooth) initial coastline submitted
to the erosion action of the sea. As discussed later
(Section~\ref{sec:smoothing}), the dynamics with different initial
morphologies can be understood from these results.

\subsection{Erosion dynamics}
\label{sec:erosion}

In the first steps of the dynamics the erosion front keeps quite
smooth and it roughens progressively as shown in
Fig.~\ref{fig1}. During the process, finite clusters are detached from
the infinite earth, creating {\em islands}. At any time, both the
islands and the coastline perimeters contribute to the damping. As the
total coastline length $L_p(t)$ increases, the sea force becomes
weaker. At a certain time step $t_f$, the weakest point of the coast
is stronger than $f(t_f)$ and the ``rapid'' dynamics stops. This
indicates that erosion reinforces the coast by preferential
elimination of its weakest elements until the coast is strong enough
to resist further erosion. Whatever the dynamics, at the stopping time
$t_f$ the coastline is irregular (see Fig.\ref{fig1}) up to a
characteristic width $\sigma$. This width $\sigma$ is defined as the
standard deviation of the final coastline depth.  More precisely,
defined $n_f$ as the mean number of points of the front lying on the
line $x$, and $x_f$ as the average position of the front, that is
\begin{equation}
x_f = \frac{\sum_{x=0}^{L_g} x
n_f(x)}{\sum_{x=0}^{L_g} n_f(x)},
\label{XFdef}
\end{equation}
then $\sigma$ is
\begin{equation}
\sigma^2= \frac{\sum_{x=0}^{L_g}(x-x_f)^2
n_f(x)}{\sum_{x=0}^{L_g} n_f(x)}.
\label{sigmadef}
\end{equation}

The (averaged) time evolution of $f(t)$ is shown in Fig.~\ref{fig3}
(left).  The dynamics depends strongly on the value of $g$. If $g$ is
large enough the dynamics is rapid and the erosion stops on an
irregular but non-fractal sea-shore (see the {\em strong coupling}
case $g=0.1$ in Fig.~\ref{fig3}).  On the opposite, if $g$ is small
enough, the dynamics last much longer and it finally stops on a
fractal sea-shore ({\em weak coupling} case). Note that the final
values of the sea-power are different from the classical percolation
threshold. This is linked to the weakening rule implemented in the
model.

It is however important to stress that the time here is a computer
step, not directly comparable with physical time. A unit of computer
step corresponds to the duration between erosion events. Such a
duration is {\em directly related to the strength or fragility of the
  coast itself}.

\subsection{Long term dynamics}
\label{sec:longtermerosion}
Of course, the real dynamics of the coasts are more complex than the
{\em ``rapid''} processes considered above. They result from the
interplay with the slow weathering processes, generally attributed to
carbonation or hydrolysis~\cite{davies2}. These processes act on
longer, geological, time scales.

\begin{figure}[h]
\centerline{\includegraphics[height=7.5cm]{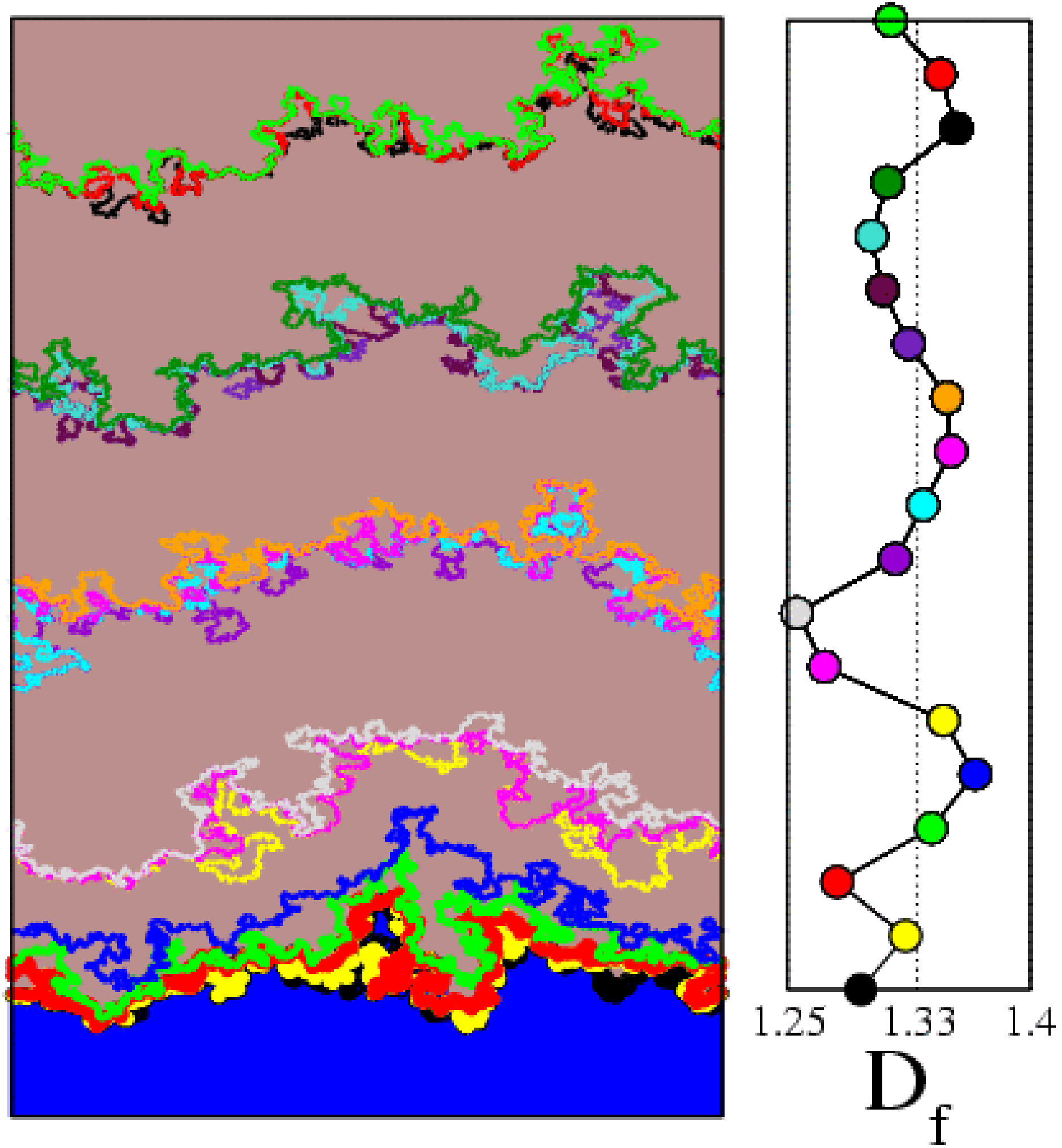}}
\caption{\label{slow} Snapshots taken during the long term erosion
dynamics for a small system ($L_0=3000$) with a moderate coupling
$g=0.002$. Color codes for successive arrest times.
Note that the measured fractal dimension fluctuates around
the universal value $4/3$ corresponding to the limit of vanishing coupling (right
panel).}
\end{figure}

In order to mimic this long term evolution after the ending of {\em
  "rapid"} erosion, the lithology parameter $l_i$ of all the coast
sites is decreased by a small fraction $\epsilon$, \emph{i.e.}
$l_i'=(1-\epsilon)l_i$ with $\epsilon\ll 1$ after the erosion has
stopped at $t_f$. One or a few coast sites then become weaker than
$f(t_f)$ and they are eroded. This exposes new sites, previously
protected, to erosion, triggering possibly a new start of the rapid
erosion dynamics up to a next arrest. This process can then be
iterated. Snapshots of the coastline at successive arrest times are
shown in Fig.~\ref{slow} together with their measured fractal
dimension.  Note that the measured fractal dimension fluctuates around
$4/3$, which is the expected fractal dimension for a very large coast
with a vanishing or very small coupling $g$ (as in Fig.~\ref{slow}).

Moreover, at each restart of erosion, a finite and strongly
fluctuating portion of the earth is eroded.  The ``slow'' weathering
mechanisms induces also small fluctuations of $f(t)$ (see
Fig.~\ref{fig3} right).  In the language of coastal
studies~\cite{haslett}, the system state evolves through a dynamical
equilibrium where small perturbations may stimulate large fluctuations
and avalanche dynamics. The coastal engineering community has recently
suggested the use of a stochastic description of the dynamics of rocky
coast erosion, characterized by episodic, discontinuous events, more
than simple constant erosion rate processes~\cite{Hall2002}. Field
inspections~\cite{Dornbusch2008} confirmed that the local variability
in rock resistances, or in general "geological contingency influences
the nature and scale of erosion processes and
thresholds"~\cite{Naylor2010}.

In our model such fluctuating dynamics is due to the underlying
criticality of percolation systems. A detailed statistical analysis of
the episodic erosion events is out of the scope of the present
paper. Nevertheless, we wish to point out that the statistics of such
events should follow power laws (similarly to what has been computed
for the etching of disordered solids at low
temperature~\cite{Kolwankar}). Interestingly, power laws have been
observed in the statistics of field coastal soft-cliff
erosion~\cite{Dong}. In~\cite{Lim2010}, using a high temporal
resolution rockfall statistics, the episodic character of the dynamics
is debated, in opposition to a continuum activity. Within our
approach, a power law statistics is expected in the framework of
percolation theory.

Note that the fluctuating dynamics of $f(t)$ in our model, are not due to fluctuations of the sea incoming power $f_0$. These could also be included, in order to mimic storms, for instance. Such fluctuations could give raise to reactivations of fast eroding events, as for the weathering mechanism. In general, we don't expect a change in the overall set of morphologies generated by the model.


\section{Fractal versus non-fractal seacoasts}
\label{sec:fractalornot}

As explained above, our model of coastal erosion shows how the
feedback between the lithology heterogeneity of shores and the damping
effect of geometric irregularity may lead, in the weak coupling case,
the coast towards a fractal geometric shape with a dimension
characteristic of percolation.  This case can be qualified as
``canonic'' since it corresponds to well established percolation
theory results.

However, under different conditions, the model may generate coastlines
with complex irregular shapes, not necessarily fractal. This happens
in the case of strong coupling morphology of "old coasts", or in the
case of transient morphologies of the weak coupling "young coasts".  A
measure that can result useful to characterize the morphology in this
case is the following.

\begin{figure}[h]
\centerline{\includegraphics[height=5cm]{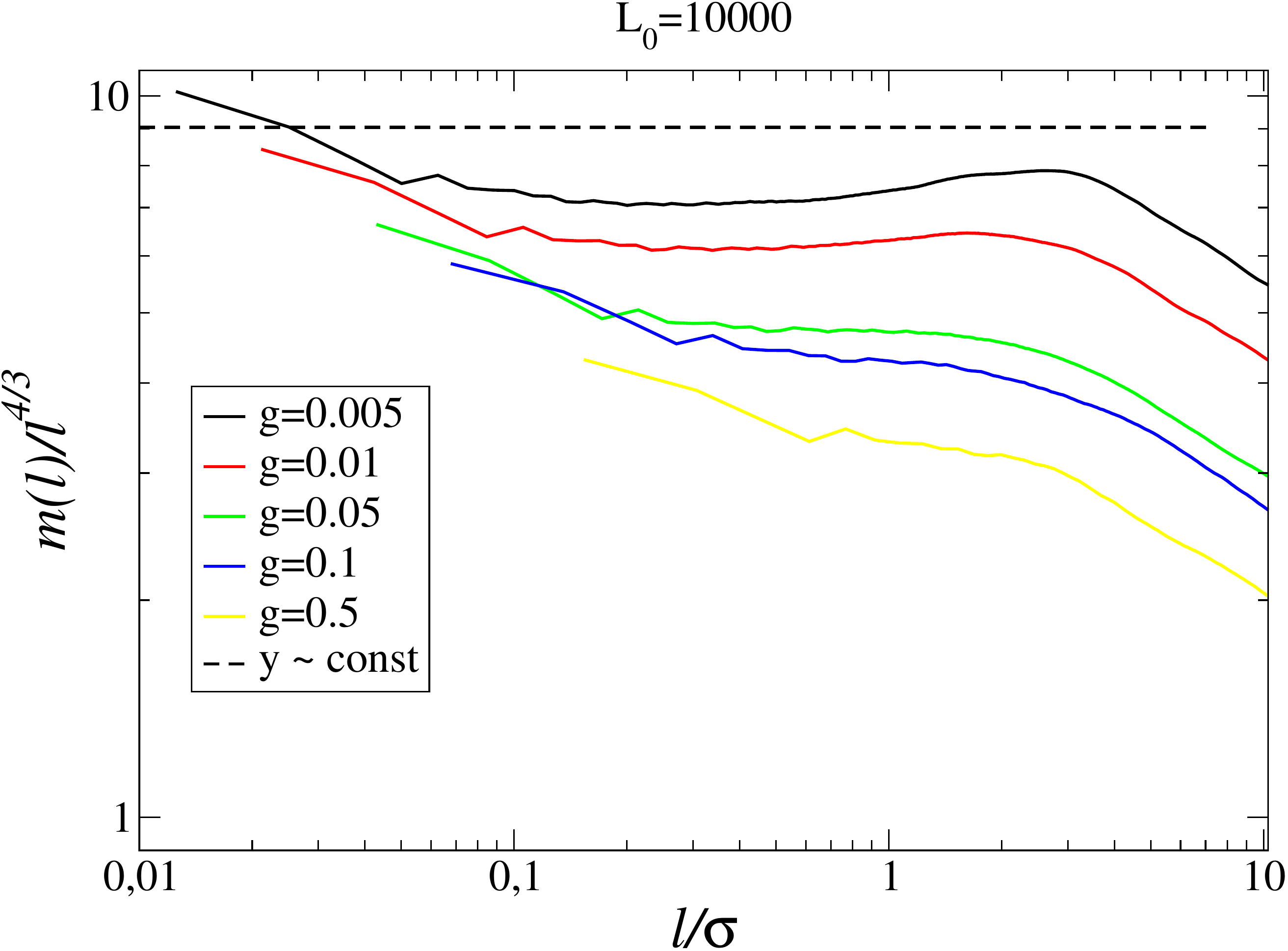}
\includegraphics[height=5cm]{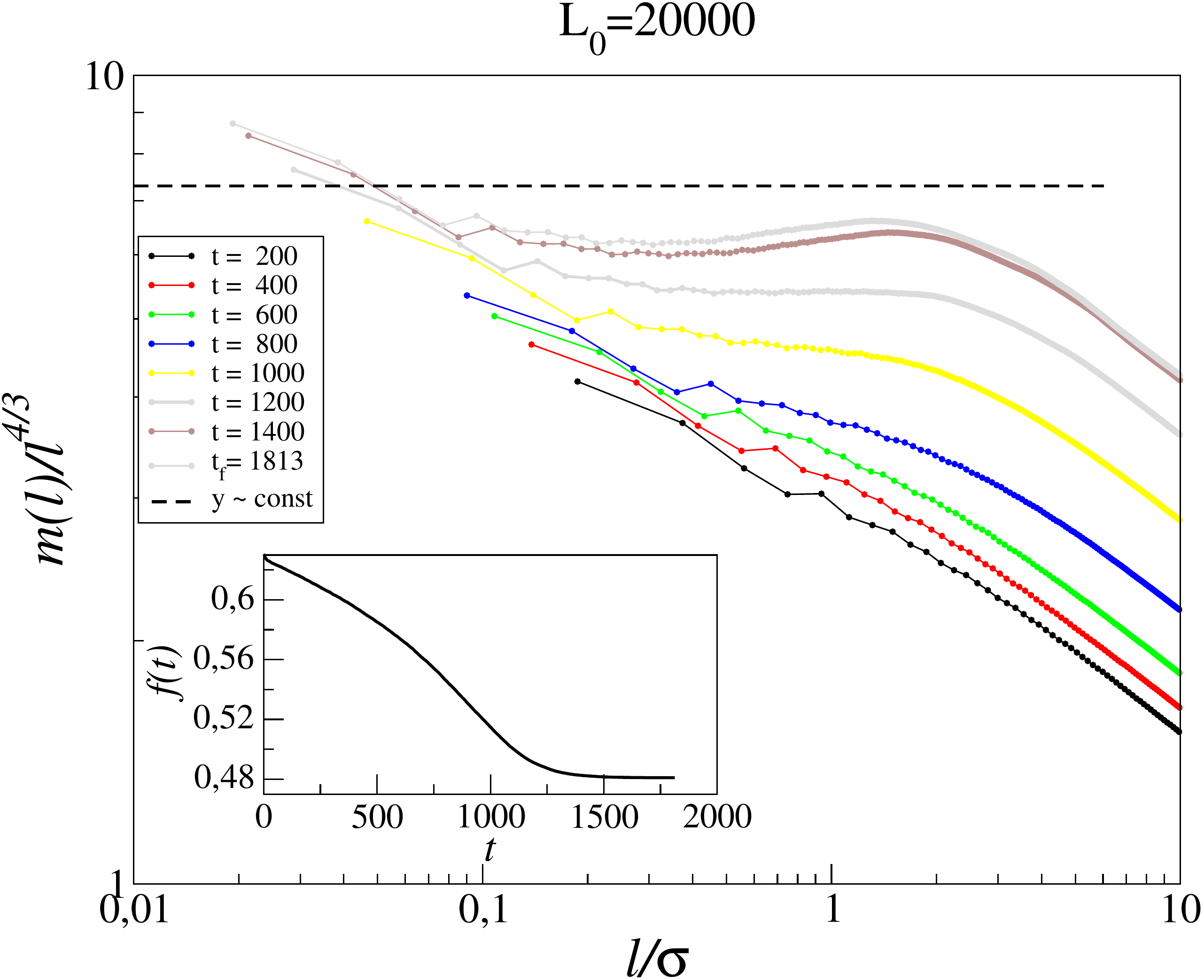}}
\caption{\label{mass-final} Left: Geometrical correlation of ``old'' coast at the end of the fast erosion dynamics. The mass plotted here
 is the local mass in a box of side $l$ centered around
 a point of the coast and averaged along this coast.
$L_0 = 10000$. \label{mass-transient} Right: Geometric correlation of the ``young'' coast
obtained during the
fast erosion dynamics. Inset: erosion strength of the sea during the
dynamics. The mass plotted here
 is the local mass in a box of side $l$ centered around
 a point of the coast and averaged along this coast. $L_0 = 20000$, $g=0.01$.
}
\end{figure}

First, one has to recall that the exponent which plays a role in the
geometry is that of the accessible perimeter, namely $4/3$.  So an
empirical way to determine if an observed coast may result of such
erosion is to measure the length of the coast contained in a square
box of side $l$. This is the classical mass method in fractal
studies. In our case, this length should be proportional to $l$ to the
power $4/3$ up to a box size of order $\sigma$. For larger boxes, the
mass should be linear as a function of $l$. This is shown in
Fig.~\ref{mass-final} for various computed final or \emph{``old''}
coasts obtained for different, but quite large, values of $g$. In this
case, as can be seen in Fig.~\ref{fig4} left, $\sigma$ is much smaller
then $100$, and the coasts appear simply rough, rather than fractal
(see Fig.~\ref{fig1}(f)).  Nevertheless, a clear flattening of the
curve, for $l\approx \sigma$ is visible.  (The fact that the signature
of the fractal exponent can be observed in a non fractal front has
been also investigated mathematically for gradient percolation
in~\cite{sapo6,sapo7}).

Interestingly, for {\em young transient} coasts as that of
Fig.~\ref{fig1}(b), which also appears rough, non fractal as in
Fig~\ref{fig1}(f), the flattening is much less evident. In
Fig.~\ref{mass-transient} we show the measure of $m(l)$ during the
fast erosion process, which eventually leads to red curve in
Fig.~\ref{mass-final}. We note that the flattening arise quite late in
the erosion process. This suggest that with respect to this measure,
transient structures behave differently than final structures.

\section{Application to more general geological situations}

Our model, as discussed until now, may appear too simplistic. Here we
discuss how our results applies more generally.

\subsection{Starting from an irregular coastline}
\label{sec:smoothing}

In the above results, the general trend is to reach a rugged
morphology starting from a flat one, an obviously artificial situation. 
Even without including specific mechanisms, mimicking a differential erosion due to convergence of sea waves (due to local topography or bathymetry, as well as specific wind directions), we can wonder what would happen in our model if one would start from an initial coast with a salient geometry.

\begin{figure}[h]
\centerline{\includegraphics[width=12cm,angle=0]{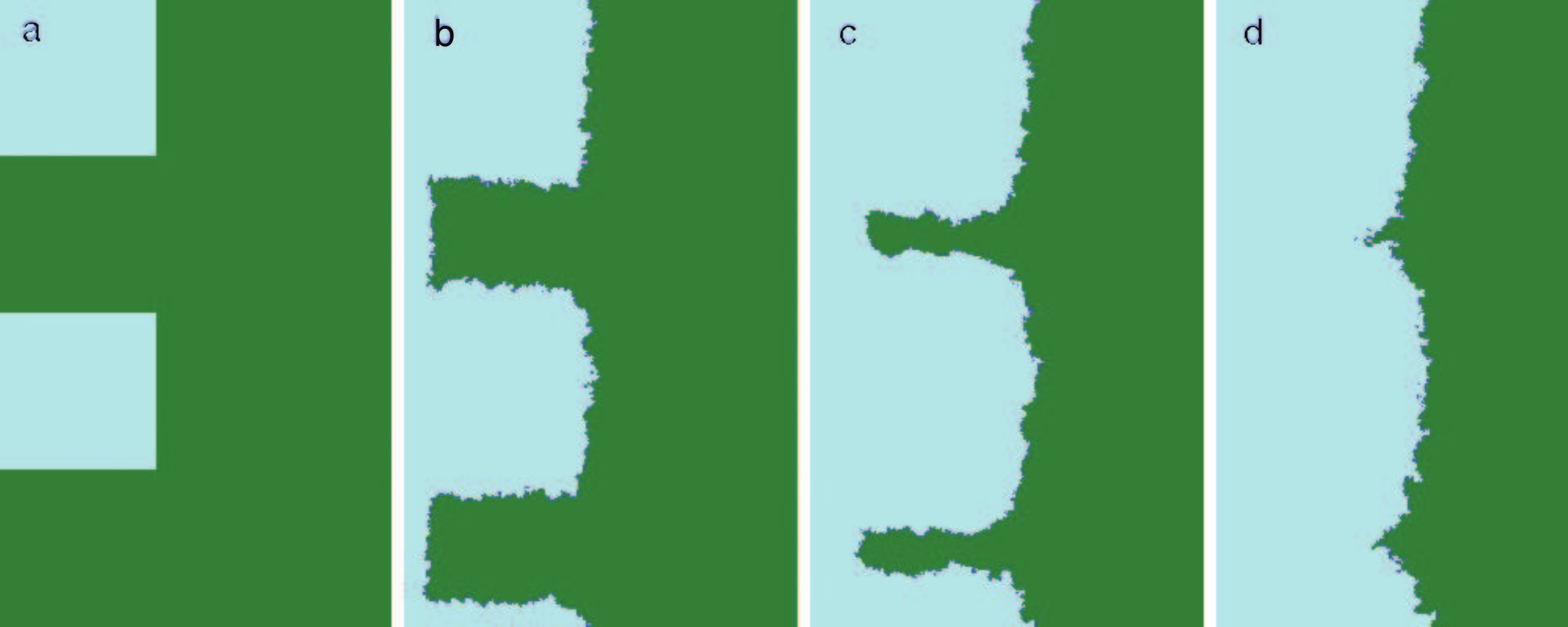}}
\caption{\label{dentedisega} Long term erosion, strong coupling starting from an irregular geometry (shore flattening): a) initial configuration; b) after $50$ erosion cycles; c) after $100$ erosion cycles; d) after $150$ erosion cycles.}
\end{figure}

\begin{figure}[h]
\centerline{\includegraphics[width=12cm]{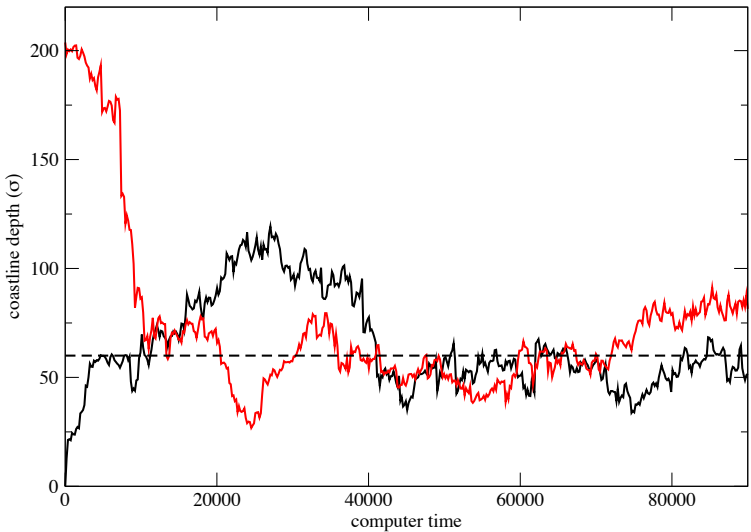}}
\caption{\label{fig-trenhaile} Depth of the coast, measured as the length scale $\sigma$ of its irregularity, defined in Eq.~\protect{\ref{sigmadef}}, as a function of computer time during the slow erosion process. Two different initial conditions are considered: large $d\equiv \sigma(0)$ (red curve) and small $d$ (black curve), compared with the average depth obtained starting from a initially flat shore $\sigma_f$ (dashed line). Both simulations are perfomed with $g=0.1$, $L=1000$ and $f(0)=0.6$. In the inset, reproduction of Fig.5 from~\protect{\cite{Trenhaile2002}}.}
\end{figure}

Let call $\sigma_f$ the typical scale length of the irregularity
produced by the dynamics starting with an initial flat coast (for
instance measured as the average of $\sigma$, defined in
Eq.~\ref{sigmadef}, on several dynamics at large times). This quantity
depends on $g$ (it is proportional to $g^{-4/7}$).  Suppose now the
case of an initial non flat coast, that is a coast with an initial
characteristic irregularity scale $d$. Then, if $d > \sigma_f$, the
erosion dynamics will initially change (decorate) the coast on the
smallest scale $\sigma_f$, keeping the larger irregularity $d$. Then,
the slow erosion dynamics will eventually loose memory of the initial
geometry, leading to a shoreline irregular up to time $\sigma_f$.
This case is shown in Fig.~\ref{dentedisega}, where a "toothed"
coastline is submitted to erosion according to our model in the case
of strong coupling (small $\sigma_f$).  Otherwise, if $d<\sigma_f$,
the erosion will increase the irregularity up to length $\sigma_f$,
which becomes the dominant irregularity scale.  In
Fig.~\ref{fig-trenhaile} we show the results of the equivalent
dynamics (same $g$ and $f(0)$) of two initially different irregular
coasts, respectively with $d<\sigma_f$ and $d>\sigma_f$. After some
time, which depends on the dynamics parameters, the depth of the coast
shows a fluctuating dynamics around $\sigma_f$, irrespectively from
the initial value of $d$. This, we think, resembles what Trenhaile had
in mind drawing a figure in his paper~\cite{Trenhaile2002} (reproduced
here as an inset in Fig.~\ref{fig-trenhaile}).  Several observation
are however in order. Here $d_{eq}$ corresponds to $\sigma_eq$, which
is related to the correlation length of the underlying percolation
process. It depends on $g$, the coupling between (geometrical) damping
and erosion. There's no need to invoke an {\em a priori} differential
erosion rate between headlands and bays. Finally, the {\em
  quasi-equilibrium regime} evoked by Trenhaile, corresponds here to
the stationary, critical dynamics, where avalanches are triggered by
slow erosion, local, events.


We think that this kind of dynamics blends two apparently contrasting
ideas: by one side the {quasi-equilibrium} dynamics predicted by
Trenhaile, on the other the image of an episodic
dynamics~\cite{Lee2001,Hall2002}. At the same time it supports the
continuum activity scenario recently proposed in~\cite{Lim2010}, where
the magnitude-frequency distribution corresponds to the avalanche
statistics of our critical dynamics, expected in our model since the
connection with percolation and observed in similar
cases~\cite{Kolwankar}.

\subsection{Geological heterogeneity}
\label{sec:geology}

It is of general consent that marine erosion processes acts on an
earth that possess its own geological identity and this obviously
influence the observed morphologies. In the above calculations and
discussion, the lithology distribution has been considered totally
random, without any spatial correlations. By this, we mean that the
lithology of neighboring "rocks" are independent random numbers. There
is {\em no correlation in the disorder}.

This is a limit case. A more realistic description should include the
concept of "geological heterogeneity". In this case the lithology
exhibits a dispersion around a local average value, which changes only
on large distances. The scale of variation of this local average is
called {\em the correlation distance} of the random lithology.

\begin{figure}[h]
\centerline{\includegraphics[height=5cm]{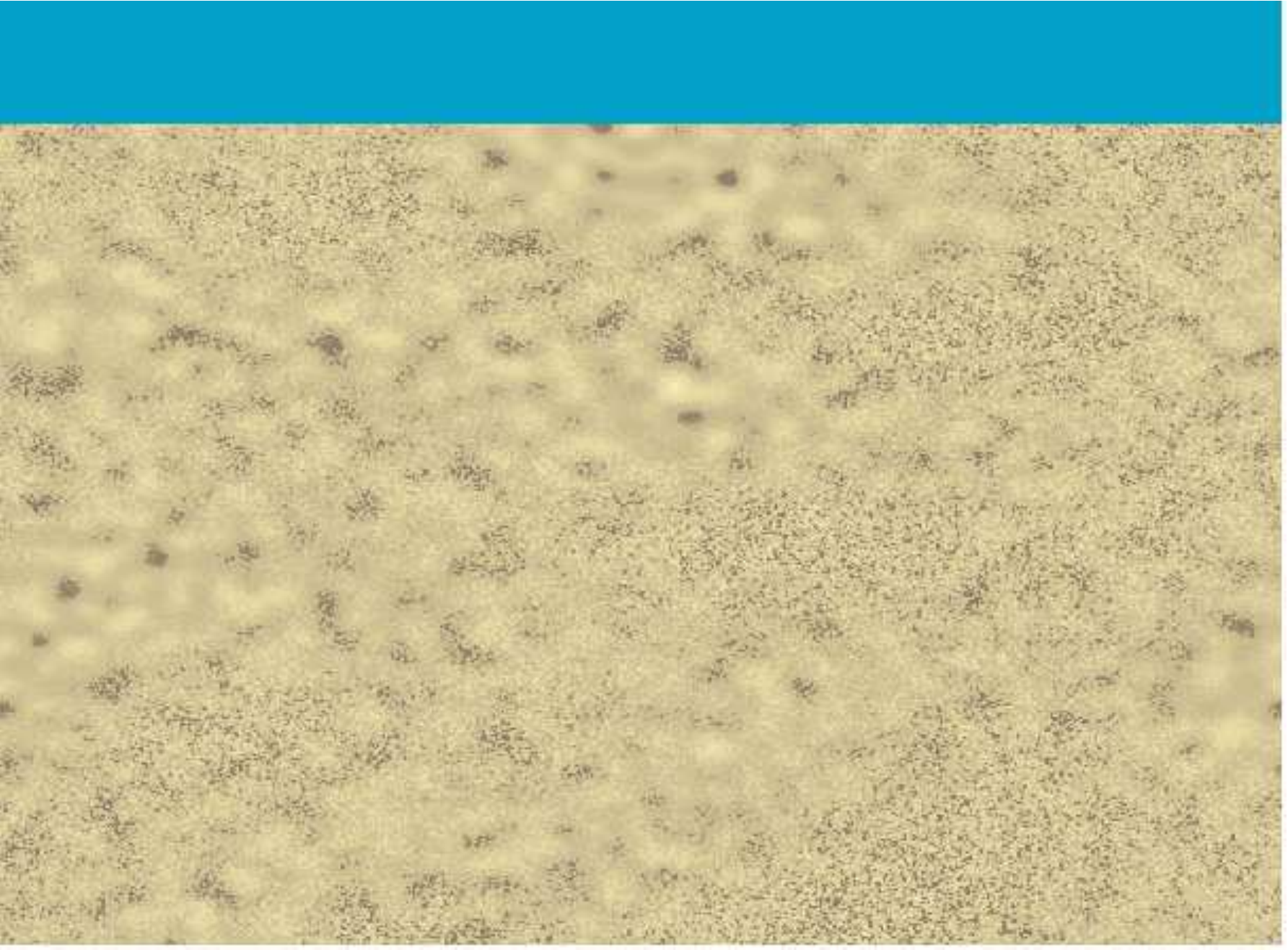}
\includegraphics[height=5cm]{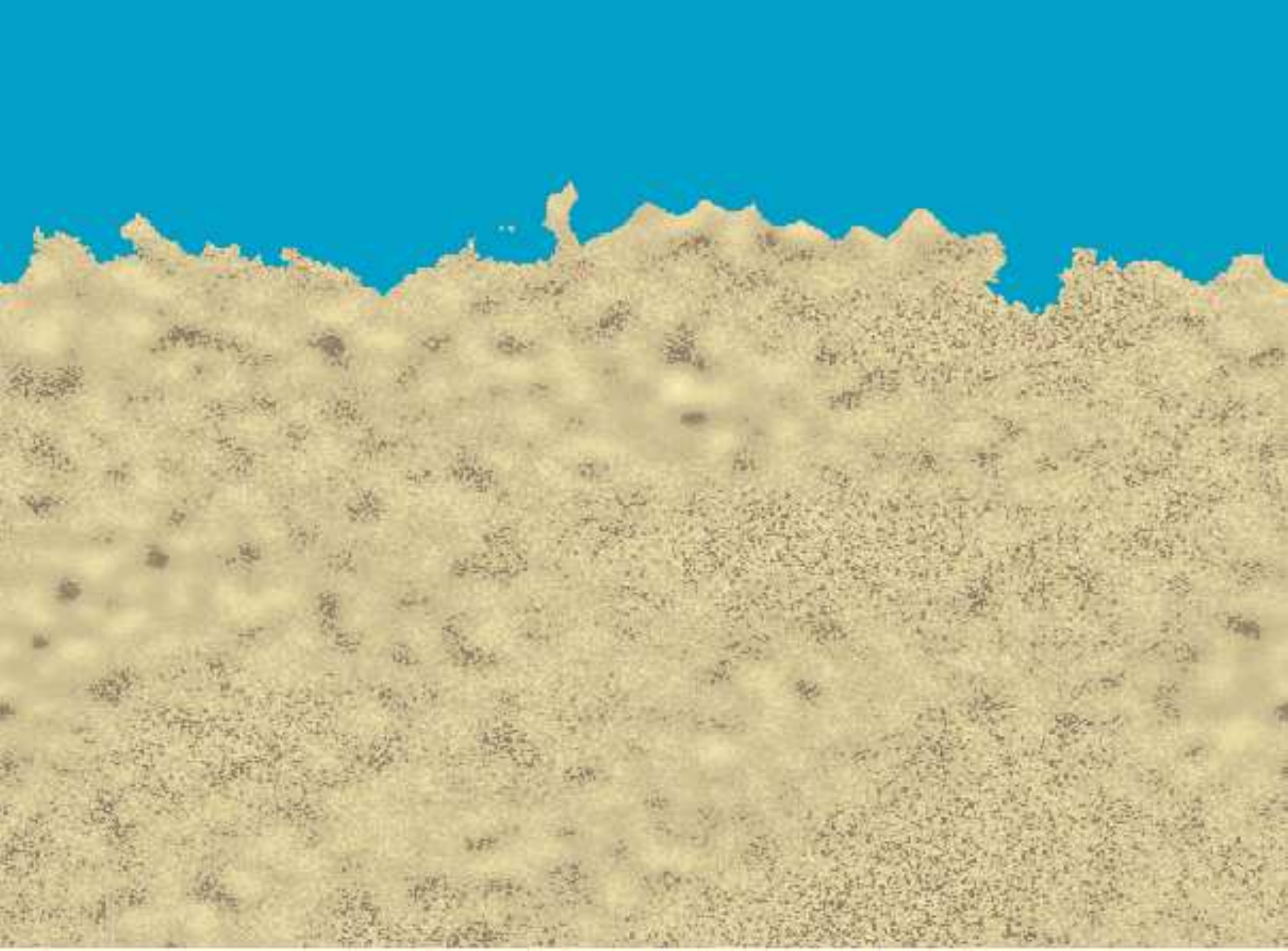}}
\centerline{\includegraphics[width=5cm,angle=90]{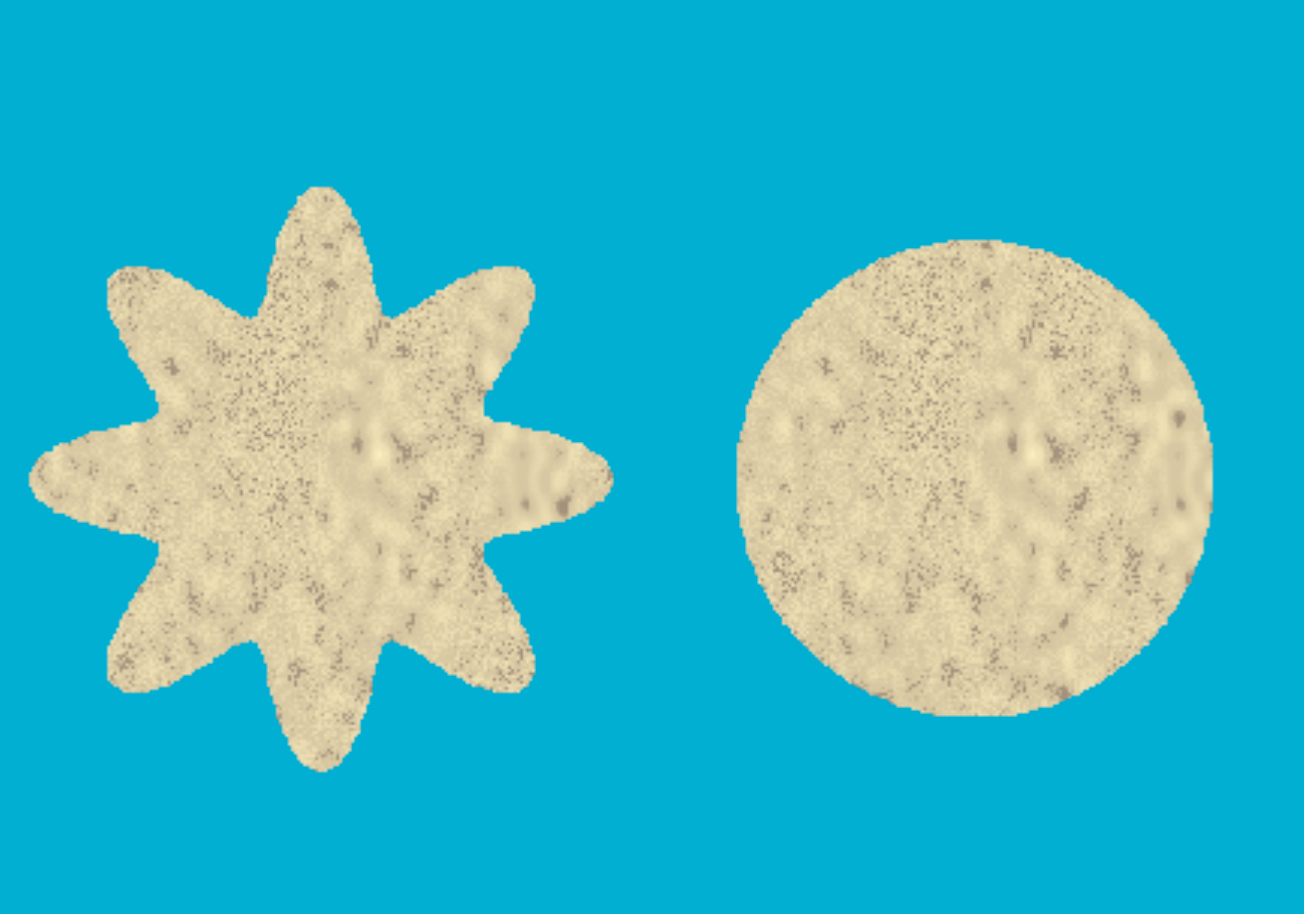}
\includegraphics[width=5cm,angle=90]{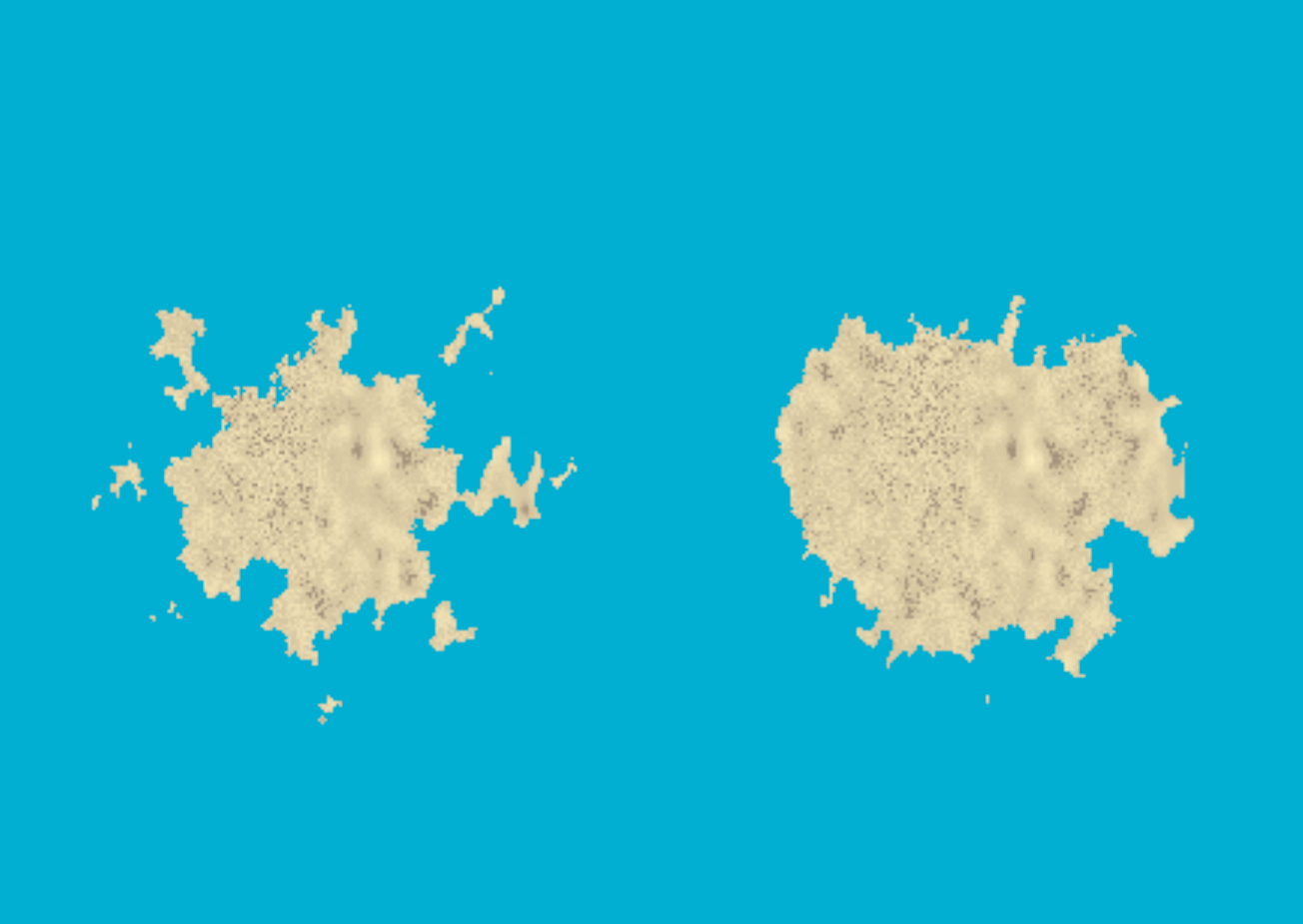}}
\caption{\label{flat-correlated} Top: Example of initial (left) and corresponding final (right) morphologies created by the erosion process in the case where the earth present regions of local correlation between lithologic mechano-chemical properties. The lithologies values are coded by light to dark beige. \label{correlated} Bottom: Example of initial (left) and corresponding
final (right) morphologies created by the erosion process with a different starting geometry. Pictures are courtesy by J.F.Colonna~\cite{colonna}.}
\end{figure}

In Fig.~\ref{flat-correlated}, the earth is constituted at the
beginning of a collection of different patches, some of which contains
long distance correlations. It then presents some regions of weak
lithology and some regions of strong lithology while other regions are
uncorrelated. The final morphology retains this heterogeneity, with
part of the shore very irregular while other regions are smoother
(similar phenomena could be invoked in order to interpret several
detailed studies of the observed self-similarity properties of
seacoasts~\cite{goodchild,andrle,bartley}).


Such irregular coastlines, which retain some strong resemblance with
real coasts, do not enter a simple fractal or scaling category. The
important fact here is that, whatever the lithological conditions,
there exists a spontaneous evolution towards irregularity that stops
spontaneously. This is again a consequence of the concept of
percolation but of course of a percolation problem asked on a
spatially correlated randomness.  Moreover, even in this case, the
initial flat coast is an idealization, and some memory of ancient
shapes could also influence the coastline morphology, as shown in
Fig.~\ref{correlated}.

\section{The role of sediments}
\label{sec:sediment}

Up to now, the results described here where obtained under the
assumption that sediments play no role, either because they do not
modify damping on the surface, either because they disappear by a
rapid transport effect.  There exists, indeed, situations in which
sediment transport can be neglected. See, for instance, the discussion
in Chapt.~18 in~\cite{davies}.

However, when sediments stay on site, they can produce two different
effects in terms of rock erosion (the role of sediments produced by
erosion as determinant for shaping sediment rich shorelines, which has
witnessed recent modeling efforts~\cite{murraynature,valvo2006}, is
not discussed here). First if they are small enough they can play the
role of little hammers accelerated by the waves. On the opposite they
can fall on the sea floor and contribute to the damping of waves. To
simplify, the small pieces increase erosion while the large heavy
pieces increase damping. The erosion increase has been neglected here
because its role should be only transitory.

Now, suppose, in an extreme scenario, that the damping is not
dominated, as was assumed above, by the interaction with the coast
perimeter but by the damping effect of the sediments. In fact, it is
known, and however poorly understood, that the shore waves are
partially damped by their interaction with the sea floor. In first
approximation, the more sediments the more damping. This means that
the equivalent quality factor would be inversely proportional to the
total amount of eroded material at time $t$ that me call $M(t)$,
rather than to the coast perimeter length. In that case the erosion
power would evolve as
\begin{equation}
f(t)=\frac{f_0}{1+\frac{g'\,M(t)}{M_0}}\,,
\label{eq3}
\end{equation}
where the factor $g'$ measures the relative contribution to damping of
accumulated sediments shore as compared with the total damping.  It
might also take care of the fraction of the sediments transported away
from the coast.  In particular erosion of very high cliffs, producing
large amounts of heavy rubble, would consequently induce a strong
coupling effect, leading to rugged but not fractal sea shores.

Note that this damping mechanism, in which the eroding force decreases
with the amount of eroded mass, leads to a one to one correspondence
with the aluminum corrosion experiments and models mentioned
above~\cite{sapo5} (where the corrosive power decreases with the
amount of corroded material). Then, the theory developed there, which
again makes connection to gradient percolation, applies also in our
case. This means that erosion will lead also to fractal or scaling
interfaces with the sames scaling geometries.

The two mechanisms may occur simultaneously, giving $f_0/f(t) = 1+g
L_p(t)/L_0+g' M(t)/M_0$, without affecting the result: more generally
we expect that any model that express some kind of link between random
erosion and increased damping would lead to the same type of
self-organized fractality or scaling.

\section{Coastline complexity}
\label{sec:complexity}

The analysis presented in Section~\ref{sec:singular}, suggests that
the coastal morphology is not the sole reflect of the inland
morphology. Before discussing specific examples in detail one should
recall that other phenomenon are known to play a role. For example
sand deposits usually smooth the irregularity of rocky coasts, filling
bays, or may display specific patterns~\cite{murraynature}. Also the
rough geometry of glacial valleys gives the very convoluted coastline
typical of fjords at large absolute latitudes. Note also that our
model has used implicitly the fact that the power giving rise to wave
excitation was uniform. Of course this may be not true on a too large
scale (think for instance to different wind or current conditions).

In this section we consider several specific locations, and we analyze
the coastline fractal morphology, in comparison with the fractal
geometry of the closest inland isolines. This is made possible by high
resolution SRTM3 data-set, which provide altimetric data of the earth
surface in a grid of $3$ arc-seconds (that is a resolution of about
$90$m) and with a vertical error smaller than $9$ meters~\cite{SRTM3}.
Thanks to this new tools, it is possible to unfold the complexity of
real coastline geometry, which is the result of the interplay of
several physical phenomena, and, consequently might go beyond the
simple model proposed here.

In the following field analysis, we restrict to coasts where there the
terrain gradient at the coast is much larger than the average gradient
in the inland, that is what we named plateau coasts, in
Section~\ref{sec:plateau}.

\begin{figure}[ht]
\centerline{\includegraphics[height=8cm]{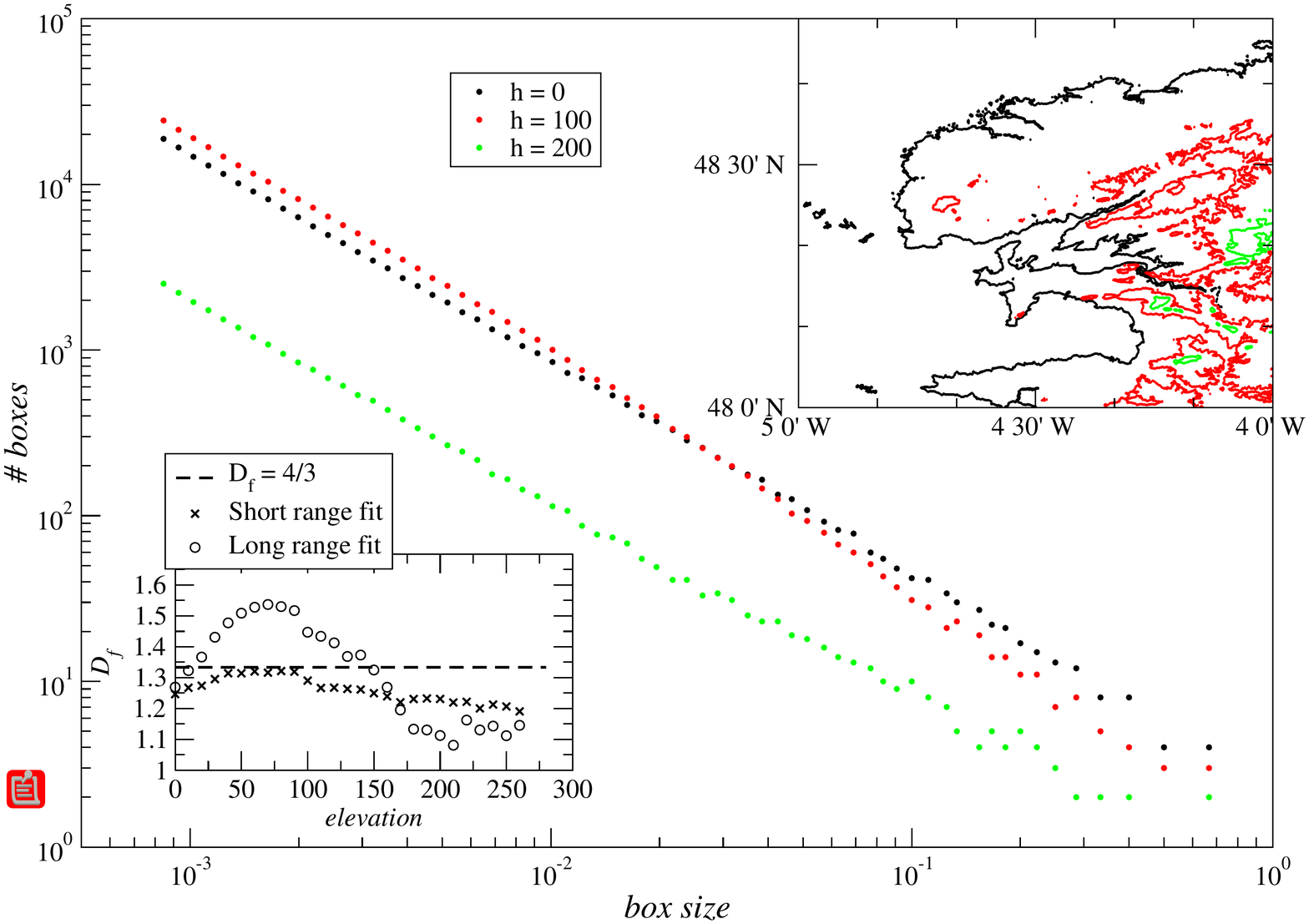}
}
\caption{\label{fig:bretagne} Analysis of Brittany coast. Scaling of box counting shows complex features, not a simple power law. This is clearly visible for the $h=100m$ isoline. Upper inset: map of the region analysed. Lower inset: slopes of box counting curves at short and long range, as a function of elevation.}
\end{figure}

An example of such coasts is the coast of Brittany, as can be seen in
Fig.~\ref{fig:bretagne}: the inland is very flat (most of the coastal
land is lower than $100$m), and the shores are usually very steep. The
coast is subject to severe storms, which can have impressive effects
on rocky cliffs~\cite{Fichaut2011}. However, the overall measured
fractal dimension of the coastline is smaller than $4/3$. This could
be due to the presence of beaches or in general sand deposits. If this
would be the case, isolines slightly higher should not be affected by
this and they should better show the effect of pure sea erosion. An
increase of fractal dimension is in fact observed for isolines
slightly higher then sea level.  However a very interesting phenomena
occurs, which can be clearly observed from the box counting curve of
the $100$m isoline: there, two appreciably different slopes appear,
one at short scale and another, steeper, for longer range. This
observation, which can not be explained by our simple model, could be
the result of different geophysical processes, acting at different
length scale and/or on different time scales. In the lower inbox of
the figure we show the fractal dimension measured restricting the
range of the fit respectively at short and long range. This effect
seems maximum at about $100$m of elevation. Interestingly, the short
range fit is very close to $4/3$ for isolines around $70$m (again note
that above $150$m the range for box-counting is too small to be
conclusive).

\begin{figure}[ht]
\centerline{
\includegraphics[height=5cm]{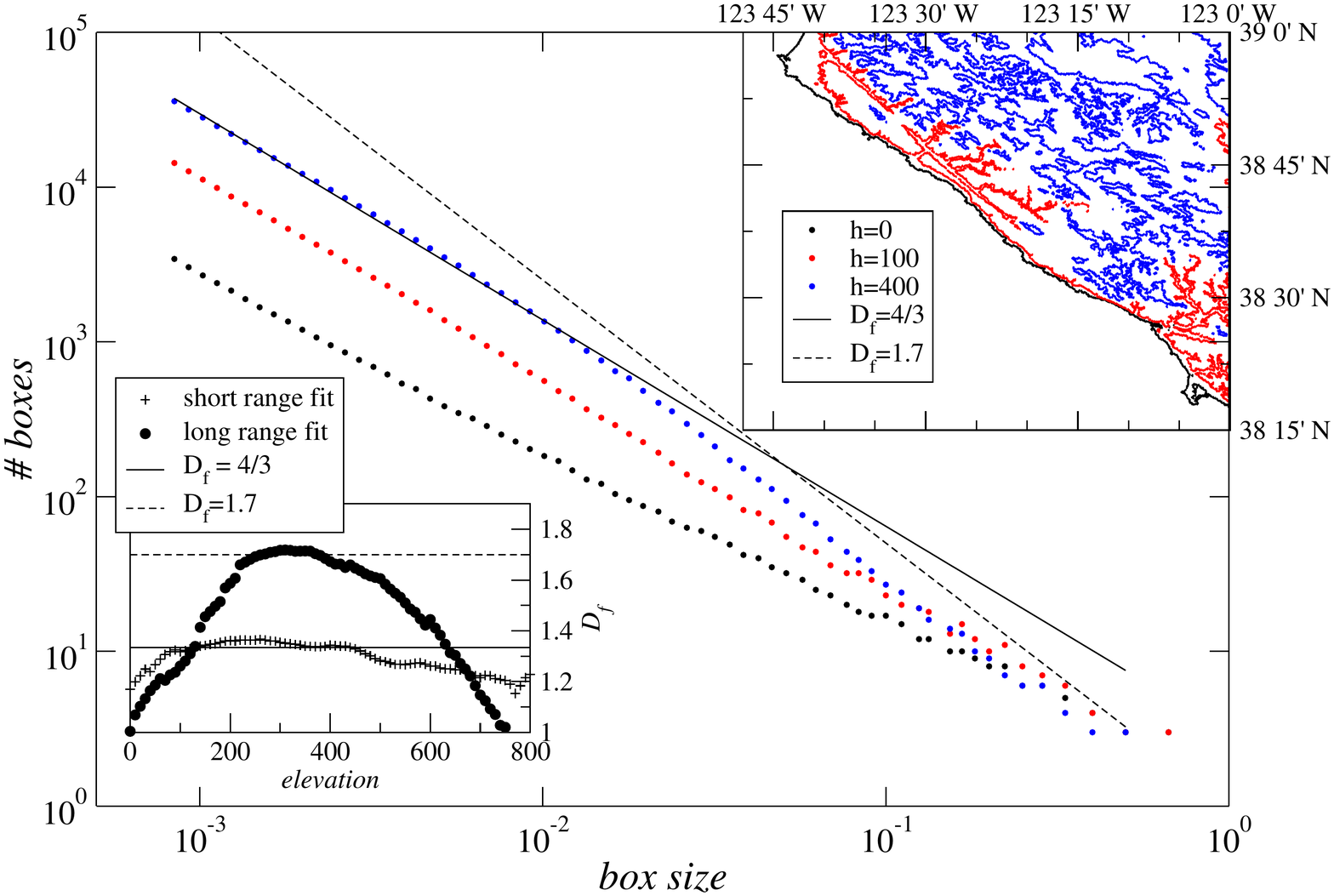}
\includegraphics[height=5cm]{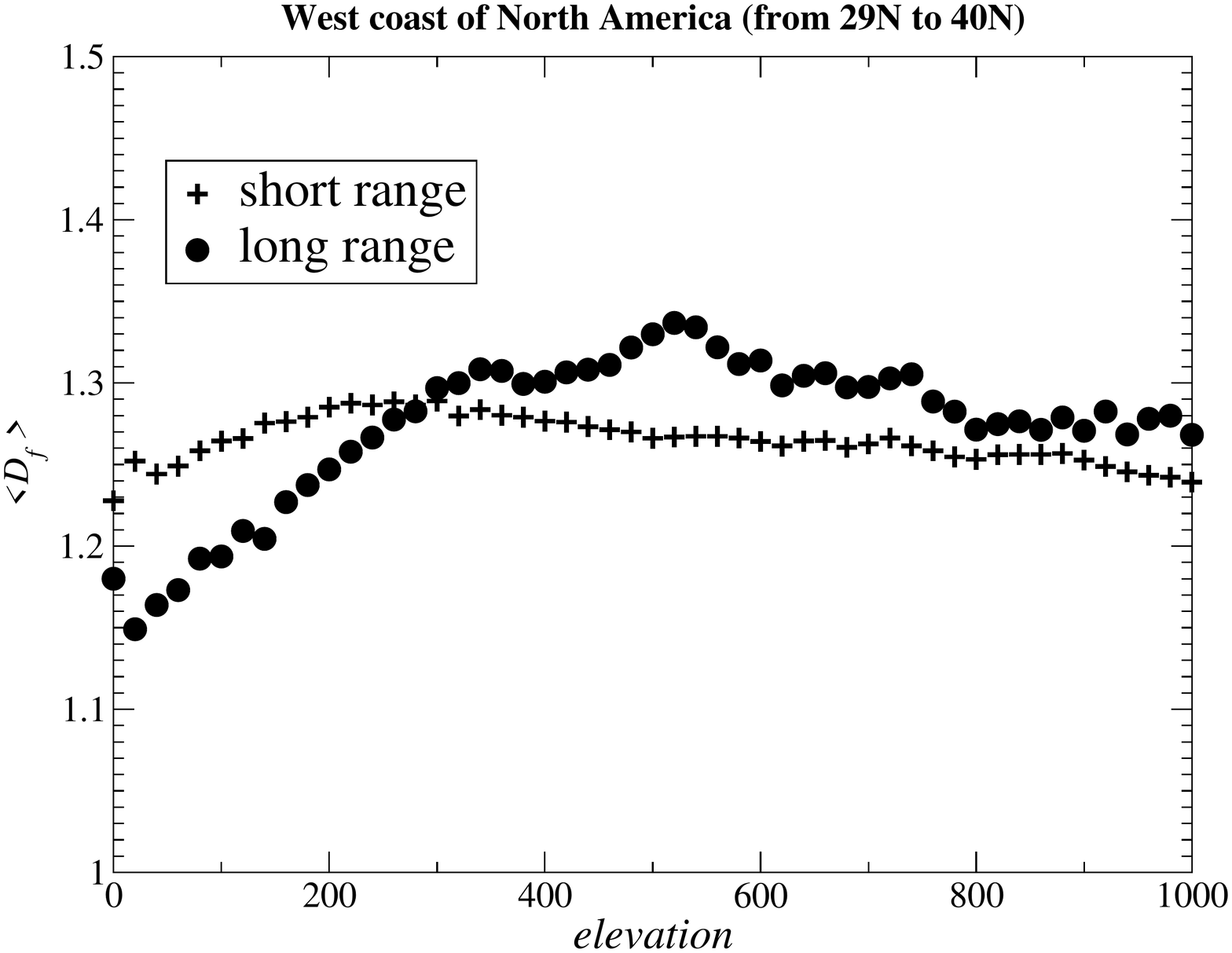}
}
\caption{\label{fig:twoslopes} Left: Analysis of northern Californian coast. Scaling of box counting shows complex features, not a simple power law. This is clearly visible for the $h=400m$ isoline. Upper inset: map of the region analysed. Lower inset: slopes of box counting curves at short and long range, as a function of elevation. Right: The average fractal dimension of West coast of north america as a function of elevation, computed at short and large range.}
\end{figure}

Such a large scale bending of isoline box-counting curves seems common
in other coastal regions, for instance in the northern coast of
California, as shown in Fig.~\ref{fig:twoslopes}. (Previous measures
of the coastline fractal dimension of American coasts are reviewed for
instance here~\cite{Jiang1998}).

In the inset of the left panel, the long range and small range slopes
are compared as a function of isoline elevation. While the short range
slope is quite constant around $4/3$ (which is quite interesting, view
the uplifting nature of this coast), at long range a wide variation is
observed.

In order to test the generality of this observation, we perform the
analysis on the whole American western coast, from $29N$ to $40N$ of
latitude, parceled in $1^o$x$1^o$ coastal cells. For each cell, we
computed the short and long range slopes as a function of isoline
elevation. In the right panel of Fig.~\ref{fig:twoslopes} we plot the
average of the measured slopes. One can see that, even if weaker, the
effect is still visible.  At the moment, we don't have any explanation
for this phenomena, even if the multifractal nature of the terrain
could possibly be invoked~\cite{lovejoy}.  A more detailed
investigation, including the eastern north American coast, which
behaves differently, will be presented in a future
publication~\cite{ustopub}.


\section{Summary and Conclusion}
\label{sec:conclusion}

In this work we have discussed a model for the formation of irregular
rocky coastal morphology. This model links the reciprocal evolution of
the erosion power with the topography of the coast submitted to that
erosion. The model reproduces at least qualitatively some of the
features of real coasts using only simple ingredients: the randomness
of the lithology and the decrease of the erosion power of the sea.

Despite the simplicity of the model, a complex phenomenology
emerges. This is not surprising in complex systems research. As stated
by Murray et al.: "The analytical lens of emergent phenomena
highlights the idea that studying the building blocks of a system –
the small-scale processes within a landscape, for example – may not be
sufficient to understand the way the system works on much larger
scales. The collective behaviors of small-scale components synthesize
into effectively new interactions that produce large-scale structures
and behaviors".~\cite{murraywhitepaper}

In our model, depending on time, on the damping strength and on
possible correlations between the lithological properties, different
coastline characteristics emerge. In the simplest case, "weak
coupling", the retro-action leads to the spontaneous formation of a
fractal seacoast with a fractal dimension $D_f = 4/3$.

The appearance of this specific fractal dimension uncovers the close,
deep connection with percolation theory and the "universality class"
of percolation in the physics of critical systems.

The dynamics in the weak-coupling case leads to a self-organized
fractality, since the fractal geometry plays the role of a
morphological attractor: whatever its initial shape, a rocky shore
will end up fractal when submitted to this type of erosion. For this
reason this case can be qualified as "canonic".

For larger coupling, and/or depending on the erosion stage, various
irregular coastlines emerge from the same model. Between these
irregular but non-fractal morphologies we have been able to
distinguish "young" or transient from "old" or final
coastlines. Future work should then include field observation, in the
hope to find such morphologies and to confront these ideas with
geologic knowledge.  Obviously, one of the first goals of field
studies will be to try to read world coastline morphology in terms of
gradient percolation in the limit of strong gradients. For these "old"
coasts, one should observe power laws with percolation exponents
\emph{although the coasts may not be fractal}~\cite{sapo6}.

At this point, one should stress that, between the enormous
geometrical variety and complexity of sea-coasts, fractal sea-coasts
and specially those with dimension close to $4/3$ should no longer be
considered anymore as "complex". On the contrary, they may in fact be
the "most simple" consequences of uncorrelated randomness and
retro-action whatever the details of retro-action. In such a frame,
they are necessary consequences of percolation phenomena. Since
percolation possess the universality properties of phase
transitions~\cite{stauffer}, the scaling properties of these coasts
should not depend on more specific processes. By this, we do not mean
that the specificity of the process no longer plays a role, for
instance on the real time scale of erosion, but that they do not
determine the global large scale geometry and scaling behaviors. More
precisely:

\begin{itemize}
\item Any retro-action damping acting on any distribution of lithology
  will lead to the spontaneous self-stabilization of an irregular
  coast. In particular, nonlinear damping effects, possibly including
  the role of turbulence, would modify the time history of erosion but
  not the scaling properties of the coast geometry.
\item Once the coast is stabilized the long term evolution will be
  stochastic. It will be triggered even by small events and will
  produce some kind of avalanche statistics. This is related to the
  fact that a stabilized sea-shore may hide "weak" lithology regions
  or more fragile patches.
\item In the field, the islands which have resisted erosion under a
  power larger than the final power $f(t_f)$, should be stronger that
  the coast itself. This could be verified on the historical data of
  known seacoasts and the evolution of neighboring islands.
\end{itemize}

In summary, the present work provides a rationale that connects
damping, as illustrated in Fig.~\ref{fig:tetrapods} left, with rocky
coast morphology, as illustrated in Fig.~\ref{fig:tetrapods} right. A
simple feedback mechanism that relates the large scale morphology with
the sea wave erosion power (usually noted with $F_W$ in the coastal
literature~\cite{sunamura92}), together with a local variability of
rock resistance ($F_R$), naturally lead to the formulation of our
minimal model, which points out how both the irregular morphology of
coastlines as well as the episodic and stochastic erosion dynamics may
both be the effect of an underlying critical (percolation) point. Of
course, such a feedback process, does not exclude the existence of
other processes acting at more local, or
meso-scales~\cite{walkden2005}.

Nevertheless, our framework confirms the idea that the final coast
emerges from a natural selection process, which eliminates the weaker
part of the coast. The resulting shoreline constitutes a strong, but
possibly fragile, barrier to sea erosion. To the extent that this idea
applies, natural coasts should be "preserved" and managed with care.

\section*{Acknowledgments}
We gratefully acknowledge illuminating and fruitful discussions with
Jens Feder and Niels Hovius, and Alistair Rowe for a reading of the manuscript.






\begin{appendices}
\section{Isoline analysis}
\label{app:isolines}

Topographic data for Earth have been obtained from the SRTM30-plus
set~\cite{srtm-plus}.  The data consists in the earth surface
elevation over a grid of points.  The resolution of the grid is $2$
minutes of degree for latitude and longitude, which corresponds, in
the region of interest, to about four kilometers. We have extracted
the coastline as the set of points at zero elevation. In fact, we use
a generalized method to extract ``isolines'' at arbitrary elevation:
to draw the isoline of level $h$, we identify on the topographic grid
all the nearest neighbor sites whose elevations, $h_1$ and $h_2$,
satisfying $h_1\geq h \geq h_2$. Using coordinates and elevations of
such points, the coordinates of the isoline point are computed via a
simple linear interpolation.

Once the isolines are found, the whole Earth surface is divided in
squares of $4$ degrees latitude x $4$ degrees longitude. In fact the
square regions are separated by only two degrees as to have an
overlapping covering of the total surface. Then we proceed in
computing the fractal dimension in each square, via the classical box
counting procedure~\cite{boxcounting}: the fractal dimension $D_f$ has
been measured through a least squares fit of the exponent of the box
counting plot in the range of $(0.06:0.6)$ degrees, which roughly
correspond to a range from few kilometers to several tenths of
kilometers. We disregard fractal dimensions computed on isoline sets
with less than $N_m=500$ points. The regression error in the fractal
dimension $D_f$ so estimated never exceeded $4\%$. The whole numerical
analysis has been repeated in the following cases: (i) full resolution
for continental land ($30$ seconds instead of $2$ minutes); (ii)
different interpolation schemes for the definitions of isolines; (iii)
different minimum number of points ($N_m=100,1000,2000$). The values
of the corresponding measured fractal dimensions are only slightly
affected, and the main results do not change.

\section{Gradient percolation}
\label{app:gp}

In order to relate the erosion model to the theory of percolation, we
need to introduce the Gradient Percolation model ({\bf
  GP})~\cite{sapo4}.  In this model, each site $(x,y)$ of the lattice
is occupied with a probability which change linearly from $1$ to $0$
in a given direction $p(x)=1-x/L_g$ ($L_g$ is the size of the lattice
in the $x$ direction).  There is then a {\em gradient} in the
occupation probability (not to be confused with the terrain gradient
in geomorphology).

In {\bf GP} there is always an infinite cluster of occupied sites as
there is a region where $p$ is larger than the standard percolation
({\bf SP}) threshold $p_c$.  There is also an infinite cluster of
empty sites as there is a region where $p$ is smaller than $p_c$.  The
object of interest is the {\bf GP} front, i.e. the external limit (or
frontier) of the infinite occupied cluster.

This front is a random fractal object with $D_f=7/4$ but the
accessible part of it is a random fractal with
$D_f=4/3$~\cite{duplantier,lawler,schramm,grossman}. It has an average
position $x_f$ and a statistical width $\sigma$ defined as
follows. For $0\leq x\leq L_g$, $n_f(x)$ is the mean number, per unit
horizontal length, of points of the front lying on the line $x$. It
measures the front density at distance $x$. The position $x_f$ and the
width $\sigma$ of the front are then defined in terms of the $n_f(x)$
by~Eq.\ref{XFdef} and~\ref{sigmadef}.

It was found in~\cite{sapo4} that the mean front is located at a
distance where the density of occupation is very close to $p_c$ or
$p(x_f) \simeq p_c$~\cite{rosso1,ziff87,quintinilla2}. It was also
found that the width $\sigma$ depends on $L_g$ through a power law
$\sigma \propto (L_g)^{\nu/(1+\nu)}$ where $\nu=4/3$ is the
correlation length exponent \cite{stauffer} so that $\sigma\propto
(L_g)^{4/7}$.  The width $\sigma$ was also shown to be a percolation
correlation length~\cite{Nolin}.

As we can see, the fractal dimension of the accessible {\bf GP} front
coincide with the value measured in the erosion model. Moreover, the
width of the front scales with respect to the gradient exactly as the
width of our coastlines do with respect to the parameter $g$,
i.e. through an exponent equal to $4/7$. See Fig. 10 right)

The reason for this is that \emph{$g$ is proportional to a gradient of
  occupation probability by the sea} from the following argument. At
time $t$, the erosion power is $f(t)$ while the sea has eroded the
earth up to an average depth $X(t)$, an increasing function of
$t$. Inverting this function, $f$ can be written as $f(t(X))$. There
exists then a spatial gradient of the occupation probability by the
sea. For small enough $g$ one can write $|df/dx | = |(g/L_0)
(dL_p(X)/dX)|$. The quantity $dL_p(X)/dX$ is a function of $g$ but to
the lowest order it is a constant independent of $g$ since even with
$g = 0$, there will be an erosion due to randomness and a consequent
perimeter evolution $L_p(t)$. Then to lowest order, the real gradient
$df/dX$ is linear in $g$, the coupling factor in the erosion model.

\end{appendices}


\begin{thebibliography}{99}

\bibitem{eric}
Eric C. F. Bird,
{\em Coasts}
(Van Nostrand Rheinhold Co., New York, 1984).
Eric C. F. Bird, M. L. Schwartz (eds),
{\em The World Coastline}
(Van Nostrand Rheinhold Co., New York, 1985).

\bibitem{penning}
E. C. Penning-Roswell, C. H. Green, P. M. Thompson, A. M. Coker, S. M. Tunstall,
C. Richards, D. J. Parker,
{\em The Economics of Coastal Management}
(Belhaven Press, London, 1992).

\bibitem{davies}
R. A. Davis, Jr, D. M. Fitzgerald,
{\em Beaches and Coasts}
(Blackwell, Oxford 2004).

\bibitem{finkl}
C.W. Finkl, Coastal classification: Systematic approaches to consider in the
development of a comprehensive
system. Journal of Coastal Research, 20(1), 166-213 (2004).

\bibitem{naylor2010} L.A.Naylor, W.J. Stephenson, A.S. Trenhaile, "Rock coast geomorphology: Recent advances and future research directions", Geomorphology {\bf 114}, 3-11 (2010).

\bibitem{Gibb1978}
Jeremy G. Gibb, Rates of coastal erosion and accretion in New Zealand, N.Z. Journal of Marine and Freshwater Research, {\bf 12} (4), pp.429-456 (1978).

\bibitem{Gibb1984}
Jeremy G. Gibb, Coastal erosion. In: Spenden I, Crozier MJ eds., Natural hazards in New Zealand, New Zealand Commission for UNESCO, (1984) pp. 134-158.


\bibitem{Kennedy2007}
David M. Kennedy, Mark E. Dickson, Cliffed coasts of New Zealand: perspectives and future directions, Journal of the Royal Society of New Zealand, {\bf 37}, 2, pp.41-57 (2007).


\bibitem{mandelbrot67}
B. B. Mandelbrot,
How long is the coast of Britain? Statistical self-similarity and fractional
dimension,
Science {\bf 156}, 636 (1967)

\bibitem{mandelbrotbook}
B. B. Mandelbrot, {\em The Fractal Geometry of Nature}
Freeman, New York (1982).

\bibitem{mandelbrot75}
B. B. Mandelbrot,
{\em Stochastic models for the Earth's relief, the shape and the fractal
dimension
of the coastlines, and the number-area rule for islands},
Proc. Nat. Acad. Sci. USA, {\bf 72}, 3825-3828 (1975).



\bibitem{geomorphologyvol91} B. Murray, and M. Fonstad, Preface: Complexity (and simplicity) in landscapes, Geomorphology, {\bf 91}, p.3-4 (2007), and the other articles in the same volume.


\bibitem{Xu1993245} T. Xu, I. D. Moore, and J. C. Gallant, 
Fractals, fractal dimensions and landscapes -- a review, Geomorphology, {\bf 8}, 245--262 (1993).



\bibitem{goodchild}
M. F. Goodchild,
Fractals and the accuracy of geographical measures.
Math. Geology, {\bf 12} 85 (1980).



\bibitem{andrle}
R. Andrle,
The West Coast of Britain: Statistical Self-Similarity vs. Characteristic Scales
in the Landscape,
Earth Surface Processes and Landforms, {\bf 21}, 955(1996) and references
therein.

\bibitem{bartley}
J. D. Bartley, R. W. Buddemeier, and D. A. Bennett,
Coastline Complexity: A Parameter for Functional Classification of Coastal
Environments,
Journal of Sea Research {\bf 46}, 87-97 (2001) and references therein.

\bibitem{sapo1}
B. Sapoval,
{\em Fractals}
(Aditech, Paris, 1989),
and
{\em Universalit\'es et fractales}
(Flammarion, Paris, 1997).

\bibitem{sapoprl}
B. Sapoval, A. Baldassarri, A. Gabrielli,
Self-stabilized Fractality of Sea-coasts through Erosion,
Phys. Rev. Lett. {\bf 93}, 098501 (2004).


\bibitem{stauffer}
D. Stauffer, A. Aharony,
{\em Introduction to Percolation Theory} (Taylor \& Francis, London, 1991).


\bibitem{fractalworld}
A. Baldassarri, M. Montuori, O. Prieto-Ballesteros, S. C. Manrubia
{\em Reading the geometry of landscapes: global topography reveals action of
geological processes on Earth}, Journal of Geophysical Research, {\bf 113}, p. E09992 (2008).


\bibitem{srtm-plus}
W. H. F. Smith, D.T T. Sandwell,
Global seafloor topography from satellite altimetry and ship depth soundings,
Science, 277, 1957-1962, (1997).
Data from {\tt http://topex.ucsd.edu/WWW\_html/srtm30\_plus.html}




\bibitem{murraynature}
A. Ashton, A. B. Murray, O. Arnault, "Formation of coastline features by large-scale instabilities induced by high-angle waves", Nature, {\bf 414}, pp.296-300 (2001).

\bibitem{davies2}
R. A. Davis, Jr,
{\em Oceanography - An Introduction to the Marine Environment},
(Wm. C. Brown Publ., Dubuque, Iowa, 1986).

\bibitem{sapo2}
B. Sapoval, O. Haeberl\'e, S. Russ,
Acoustic properties of irregular and fractal cavities,
J. Acoust. Soc. Am., {\bf 102}, 2014 (1997).

\bibitem{sapo3}
B. H\'ebert, B. Sapoval, S.Russ,
Experimental study of a fractal acoustic cavity,
J. Acoust. Soc. Am., {\bf 105}, 1567 (1999).

\bibitem{simon}
S. Felix, M. Asch, M. Filoche, and B. Sapoval,
Localization and Increased Damping due to Localization in Irregular Acoustical
Cavities,
Journal of Sound and Vibration, 299 965-976 (2007).

\bibitem{simon2} S. Felix, B. Sapoval, M. Filoche, and M. Asch, Enhanced wave absorption through irregular interfaces, Eur. Phys. Lett.,{\bf85}, 14003 (2009).


\bibitem{patent}French patent 03/00881, US patent 7308965B2.


\bibitem{balasz}
L. Balasz,
Corrosion front roughening in two-dimensional pitting of aluminum thin layers,
Phys. Rev. E, {\bf 54}, 1183-1189 (1996).




\bibitem{sapo5}
B. Sapoval, S. B. Santra, Ph. Barboux,
Stable Fractal Interfaces in the Etching of Random Systems,
Europhys. Lett., {\bf 41}, 297 (1998),
and,
A. Gabrielli, A. Baldassarri, B. Sapoval,
Surface Hardening and Self-Organized Fractality Through Etching of Random
Solids,
Phys. Rev. E,  {\bf 62}, 3103-3115, (2000).


\bibitem{Trenhaile2002}
A. S. Trenhaile, Rock coasts , with particular emphasis on shore platforms, Geomorphology, {\bf 48}, 7-22 (2002).



\bibitem{manual}
{\em Shore protection manual}
(Dept. of the Army Waterways Exp. Station, Vicksburg, Mississippi, {\bf 2},
1984).





\bibitem{duplantier}
Duplantier, B. (2000), Conformally invariant fractals and potential theory,
Phys. Rev. Lett., 84, 1363.

\bibitem{lawler}
Lawler, G. F., O. Schramm, and W. Werner (2004), On the scaling limit of
planar self-avoiding walk, Proc. Symp. Pure Math., 72, 339.

\bibitem{schramm}
Schramm, O. (2006), Conformally invariant scaling limits (an overview and
a collection of problems), in Proceedings of the International Congress of
Mathematicians, Madrid, August 22- 30, 2006, edited by M. Sanz-Sole
et al., Eur. Math. Soc., Zurich, Switzerland. (Available at http://arxiv.org/
abs/math/0602151.)

\bibitem{grossman}
T. Grossman, A. Aharony,
Accessible external perimeters of percolation clusters,
J. Phys. A, {\bf 20}, L1193-1201 (1987).

\bibitem{sapo4}
B. Sapoval, M. Rosso, J. F. Gouyet,
The fractal nature of a diffusion front and relation to percolation.,
J. Phys. Lett. (Paris), {\bf 46}, L149 (1985).


\bibitem{rosso1}
M. Rosso, J.-F. Gouyet, B. Sapoval,
Determination of percolation probability from the use of a concentration
gradient,
Phys. Rev. B {\bf32}, 6035 (1985).

\bibitem{sapoval-intro} B. Sapoval, M. Rosso, and J.F. Gouyet, Fractal Interfaces in Diffusion, Invasion and Corrosion, in The Fractal Approach to Heterogeneous Chemistry (D. Avnir ed.), p.227, John Wiley and Sons (1989).


\bibitem{boffetta}
G. Boffetta, A. Celani, D. Dezzani and A. Seminara,
How winding is the coast of Britain? Conformal invariance of rocky shorelines,
Geophysical Research Letters 35, L03615 (2008).



\bibitem{haslett}
S. K. Haslett,
{\em Coastal Systems}
(Routledge, Taylor and Francis, London, 2000).



\bibitem{Hall2002}
J. W. Hall, I. C. Meadowcroft, E. M. Lee, and P. H. A. J. M. van Gelder, Stochastic simulation of episodic soft coastal cliff recession. Coastal Enginereering {\bf 46}, 159-174 (2002).


\bibitem{Dornbusch2008} U. Dornbusch, D. A. Robinson, C. A. Moses, and R. B. G. Williams,  Temporal and spatial variations of chalk cliff retreat in East Sussex, 1873 to 2001, Marine Geology, {\bf 249}, 271–282 (2008). 

\bibitem{Naylor2010}
L. A. Naylor, and W. J. Stephenson, On the role of discontinuities in mediating shore platform erosion, Geomorphology, {\bf 114}, 89-100 (2010).


\bibitem{Kolwankar}
K.M. Kolwankar, M. Plapp and B. Sapoval,
Percolation dependent reaction time in the etching of disordered solids
Eur. Phys. Lett., {\bf62},519(2003).



\bibitem{Dong}
P. Dong and F. Guzzetti,
Frequency-size statistics of coastal soft-cliff erosion,
J. Watertw., Port, Coastal, Ocean Eng.,{\bf 131},37(2005).


\bibitem{Lim2010}
M. Lim, N. J. Rosser, A. J. Robert, D. N. Petley, 
Erosional processes in the hard rock coastal cliffs at Staithes, North Yorkshire, Geomorphology, {\bf 114}, 12-21 (2010).

\bibitem{sapo6}
A. Desolneux, B. Sapoval, A. Baldassarri,
Self-Organised Percolation Power Laws with and without fractal geometry in the
etching of random solids,
in Fractal Geometry and Applications: A jubilee of Benoit Mandelbrot
(M. L. Lapidus and M. van Frankenhuijsen, eds.)  Proc. Symposia Pure Math.,
vol. 72, Part 2, pp. 485-505 (2004).

\bibitem{sapo7}
A. Desolneux, B. Sapoval,
Percolation fractal exponents without fractals and a new conservation law in
diffusion,
Europhys. Lett. {\bf72}, 997-1003 (2005).


\bibitem{Lee2001}
E. M. Lee, J. W. Hall, and I. C. Meadowcroft, Coastal cliff recession: the use of probabilistic prediction methods, Geomorphology, {\bf 40}, 253 - 269 (2001).


\bibitem{valvo2006} L. M. Valvo, A. B. Murray, A. Ashton, "How does underlying geology affects coastline change? An initial modeling investigation", Journal of Geophysical Research, {\bf 111}, p.F02025 (2006).


\bibitem{SRTM3} T. G. Farr,  et al., {\em The Shuttle Radar Topography Mission}, Rev. Geophys., {\bf 45}, RG2004 (2007).


\bibitem{Fichaut2011} B. Fichaut, and S.Suanez, Quarrying, transport and deposition of cliff-top storm deposits during extreme events: Banneg Island, Brittany, Marine Geology, {\bf 283}, 36--55 (2011).



\bibitem{Jiang1998} J. Jiang, and R. Plotnick, Fractal Analysis of the Complexity of United States Coastlines, Mathematical Geology, {\bf 30}, 535--546 (1998).







\bibitem{colonna}
\texttt{http://www.lactamme.polytechnique.fr/}

\bibitem{lovejoy} J.S. Gagnon, S. Lovejoy, D. Schertzer, {\em Multifractal earth topography}, Nonlin. Processes Geophys., {\bf 13}, p.541-570 (2006).

\bibitem{ustopub} A.~Baldassarri, B.~Sapoval, in preparation.


\bibitem{murraywhitepaper} B. Murray, et al., Geomorphology,
  Geomorphology, complexity, and the emerging science of the Earth's
  surface, {\bf 103}, 496--505 (2009)

\bibitem{sunamura92} T. Sunamura, The Geomorphology of Rocky Coasts, John Wiley and Sons, Chichester, UK, ISBN: 0 471 91775 3 (1992).

\bibitem{walkden2005} M. J. A. Walkden, and J. W. Hall, A predictive Mesoscale model of the erosion and profile development of soft rock shores. Coastal Engineering. {\bf 52}, 535-563 (2005).



\bibitem{boxcounting} K. Falconer, {\em Fractal geometry mathematical foundations and applications} (Wiley, New York, 1990)



\bibitem{ziff87}
R. M. Ziff, B. Sapoval,
{\em The efficient determination of the percolation threshold by a frontier-generating walk in a gradient},
J. Phys. A Math. Gen. {\bf19}, L1169-1172 (1986).

\bibitem{quintinilla2}
J. Quintanilla,
{\em Measurement of the percolation threshold for fully penetrable disks of different radii}, Phys. Rev. E {\bf63}, 061108 (2001).


\bibitem{Nolin}
Pierre Nolin, {\em Critical exponents of planar gradient percolation}, Annals of Probability {\bf 36},  pp.1748-1776 (2008)


\end{thebibliography}
\end{document}